\documentclass[useAMS,usenatbib]{mn2e}
\usepackage{epsfig}
\usepackage{times}

\newcommand{\kpc}{\>{\rm kpc}} 
\newcommand{\Mpc}{\>{\rm Mpc}}
\newcommand{\kms}{\>{\rm km}\,{\rm s}^{-1}} 
\newcommand{\Msun}{\>{\rm M_{\odot}}}
\newcommand{\bm}[1]{\mathbf{#1}}
\def\be{\begin{eqnarray}} \def\ee{\end{eqnarray}}
\def\ben{\begin{eqnarray*}} \def\een{\end{eqnarray*}}
\def\se#1{\S\ref{sec:#1}} \def\fig#1{Fig.~\ref{fig:#1}}
\def\equ#1{Eq.~(\ref{eq:#1})} \def\tab#1{Table.~\ref{tb:#1}}

\title{ Multi-Dimensional Density Estimation and
  Phase Space Structure Of Dark Matter Halos}

\author[S. Sharma, M. Steinmetz] 
{Sanjib Sharma$^{1,2}$\thanks{E-mail:sharma@physics.arizona.edu}, Matthias Steinmetz$^{2}$\thanks{E-mail:msteinmetz@aip.de} \\
$^1$Department of Physics, University of  Arizona, Tucson, 85721 U.S.A\\
$^2$Astrophysikalisches Institut Potsdam, An der Sternwarte 16, 14482 Potsdam, Germany}

\begin{document}
\maketitle
\label{firstpage}

\begin{abstract}
We present a method to numerically estimate the densities of a
discretely sampled data based on a 
binary space partitioning tree.  
We start with a root node containing all the particles and then
recursively divide each node into two nodes each containing roughly 
equal number of particles,until each of the nodes 
contains only one particle. The volume of such a leaf node provides an
estimate of the local density and its shape provides an estimate of the variance.
We implement an entropy-based node splitting criterion that results in a 
significant improvement in the estimation of densities compared to
earlier work.  The method is completely metric free and
can be applied to arbitrary number of dimensions.
We use this method to determine the appropriate metric at each 
point in space and then use kernel based methods for calculating the
density. The kernel smoothed estimates were found to be  more 
accurate and have lower dispersion. We apply this
method to determine the phase space densities of dark matter halos 
obtained from cosmological N-body
simulations. We find that contrary to earlier studies, the volume distribution function $v(f)$ of
phase space density $f$ does not have a constant slope but rather a
small hump at high phase space densities. We demonstrate that a model in which a
halo is made up by a superposition of Hernquist spheres is not capable in
explaining the shape of $v(f)$ vs $f$ relation, whereas a model which takes into 
account the contribution of the main halo separately roughly reproduces 
the behavior as seen in simulations. The use of the presented method is
not limited to calculation of phase space densities, but can be
used as a general-purpose data-mining tool and  due to its speed 
and accuracy it is ideally suited for analysis of large
multidimensional data sets.
\end{abstract}


\section{Introduction} \label{sec:intro}
One of the basic problems in data mining is to estimate the probability distributions or density distributions
based on a discrete set of points (particles) distributed in a multidimensional space.
Once the density distribution is known expectation values of 
other quantities of interest can be derived.
Considering the huge amounts of data both astronomy and other fields
are facing there is a need for methods that are accurate
flexible and fast. However, most of the 
existing methods encounter problems when applied to higher dimensions.
In the particular application of N-body simulations, the estimate of phase space densities is one such
problem as it  requires an efficient and flexible method for 6 dimensional phase
space density estimation for a large variety of equilibrium and 
non-equilibrium solutions of largely different topology 
(e.g. highly flattened disks, spheroidal but anisotropic halos, spheroidal nearly isotropic ellipticals).

The simplest method for density estimation is the $k$ nearest neighbor.
Consider the radius $r$ enclosing $k$ nearest neighbors then density is
given by $k/V_{d}(r)$ where $V_{d}(r)$ is the volume enclosed by 
a $d$-dimensional sphere of radius $r$ \citep{lq65}. A more accurate method than
this is the kernel density estimation (KDE) or popularly known as SPH 
\citep{1977MNRAS.181..375G,1977AJ.....82.1013L,1986desd.book.....S}. 
The results are sensitive to the choice of kernel function and the bandwidth of the kernel or in other 
words the number of smoothing neighbors. The later being more important. 
Variable bandwidth estimators are more superior as compared to the
fixed bandwidth estimators. For the multidimensional case 
simple isotropic bandwidths perform poorly when the data has 
an anisotropic distribution. In this case one needs to 
select different bandwidths in different dimensions. 
In general a co-variance matrix is determined and the bandwidth 
is selected so as to have constant co-variance in all directions. 
This leads to anisotropic kernels.
The Delaunay tessellation \citep{1992stca.book.....O,2000stca.conf.....O,2000A&A...363L..29S,1996MNRAS.279..693B}
which tessellates space into disjoint regions, performs much better 
for anisotropic data.  Delaunay tessellation is very
accurate but also very time consuming.

Most existing methods, including both KDE  and Delaunay Tessellation
require an a-priori definition of a metric of the n-dimensional space under
investigation. A suboptimal choice of metric results in a poor
estimate of the density.
Metric based density estimators provide optimal approximations, only if 
co-variance of the data is identical along all dimensions, locally at
each point in space. In general, however, data is  non-homogeneous and 
anisotropic. Consequently the above conditions cannot be realized 
by assuming a global scaling relation among different dimensions.
A method is required that is adaptive to the data under investigation.
Recently a new method dubbed FiEstAS which is metric free has been proposed by
\citet{asca05}. FiEstAS is also very fast
and efficient. 
The method relies on a repeated binary decomposition of space
(organized by a tree data structure) 
until each volume element contains exactly one particle.
The accuracy of the method depends upon the criteria used for
splitting the nodes. In the simplest implementation the
dimension to be divided is chosen either randomly or alternately,
guaranteeing equal number of divisions for each dimension. 
The more a particular dimension is tessellated the higher 
the resolution achieved in that dimension. Ideally we need a 
scheme which makes more divisions in the dimension along which there
is maximum variation and few divisions (or none) along
which there is minimum (or no) variation. 
However, the scheme as described above is data blind and thus fails to optimize the 
number of divisions to be made in a particular dimension.

In this paper we propose and evaluate a splitting criterion that is based upon the concepts of
{\it Information Theory} \citep{Shan48,Shan49,Kay03,Neil99}. 
Space is tessellated along the dimension having the minimum
entropy ({\it Shannon Entropy}) or in other words maximum information .
Consequently, this scheme optimizes the number of divisions to be made in a
particular dimension so as to extract maximum information from the data.
This method can also be used to determine the metric that
locally gives approximately constant covariance. Kernel based methods can then
be used to estimate the densities.

As an application we study the phase space density of dark
matter halos obtained from cosmological simulations. 
The code is available upon request and in future we plan to make it
publicly available at the following url \\
https://sourceforge.net/projects/enbid/

\section{Algorithm}
The basic problem is to estimate the density function $\rho(\bm{x})$  
from a finite number $N$ of data points $\bm{x^1,x^2..x^N}$ drawn from
that density function. Here $\bm{x^{i}}$ is a vector in a space of $d$
dimensions having components $x^{i}_{1},x^{i}_{2} .. x^{i}_{d}$.
The overall procedure of our algorithm EnBiD ({\bf En}tropy Based {\bf
  Bi}nary
{\bf D}ecomposition) consists of 
three steps, which we will describe in detail below.
First we tessellate the space into mutually disjoint hypercubes each
containing exactly one particle. If $V^{i}$ is the volume of the
hypercube containing $i$th particle then its density is $m_i/V^{i}$.
Second we apply the boundary corrections to take into account
the arbitrary shape of the volume containing the data. Third we apply a
smoothing technique in order to reduce the noise in the density estimate.

\subsection{Tessellation}
We start with a root node containing all particles. The node is 
divided by means of a hyper plane perpendicular to one of the axis into
two nodes each containing half the particles. If $j$ is the
dimension along which the split is to be performed, the position of
the hyper plane is given by the median of
$x_{j}$. The process is repeated recursively
till each sub-node contains exactly one particle(so-called leaf
nodes). Let $V_i$ be
the volume of the leaf node containing particle $i$, and $m_i$ be the
particle mass, then the density is given by $\rho_i = m_i/V_{i}$.
An alternative to this, as was originally done in FiEstAS,
is to calculate the mean $<x_j>$ and then
identify two points one on each side which are closest to the mean.
The split point is then chosen midway between these two points. 
$x_{cut}=(x_{left}+x_{right})/2$. This speeds up the tessellation.

In the implementation of FiEstAS the splitting axis
alternates between the considered dimensions, which guarantees roughly
equal number of divisions per dimension. 
In the calculation of phase space densities the real and velocity space are
known to be Euclidean. Therefore the splitting is done alternately in real and
velocity space and in each subspace the axis with highest elongation
($<x_j^2>-<x_j>^2$) is chosen to be split. This generates cells that are
cubical rather than elongated rectangular in the aforementioned subspace, 
and also helps alleviate numerical problems that arise when two
points have very close values of a particular co-ordinate.
We call this decomposition which is implemented in FiEstAS as  {\it Cubic
 Cells} while the one free from this as {\it General}.

For $N$ particles the binary decomposition  results in $2N-1$ nodes
out of which there are $N$ leaf nodes each having one particle.
The more a particular dimension is tessellated, the more the
resolution in that dimension. However, for data that is uniformly
distributed in a particular dimension there is actually no need to
perform a split in that dimension. This fact can be exploited to increase
the accuracy of the results.

For each node we calculate the Shannon entropy $S_j$ along each dimension (or subspace) and then select
the axis (subspace) with minimum entropy. 
The dimension having minimum entropy guarantees
maximum density variation or clustered structures in that dimension. 
In other words we split the dimension that has the maximum amount of information. 
The entropy $S$ along any dimension or subspace is estimated by
dividing the dimension or subspace into $N_b$ bins of equal size and calculating 
the number of points $n_i$ in each bin (we choose $N_b$ to be equal to the number 
of particles in each node). The probability that a
particle is in the $i$th bin is given by $p_i=n_i/N$ where $N$ is the
total number of particles. The entropy is then given by
\begin{equation}
S=-\sum_{i=1}^{n} p_i \log(p_i)
\end{equation}

Rather than treating each dimension independently it is also possible
to select a subspace (real or velocity space) with minimum entropy and then 
choose an axis with maximum elongation from this subspace ({\it Cubic Cells}).
This provides slightly lower dispersion in estimated densities.

\subsection{Boundary correction}
The data in general might have an irregular shape and  may not be 
distributed throughout the rectangular volume of the root
node. Consequently, the densities of particles near the
boundary can be underestimated. 
This is not an issue for systems with periodic
boundary conditions but it would be for systems which are for example spherical. 
In higher dimensions this correction becomes even more
important since the fraction of particles that lie near the boundary
increases sharply with number of dimensions\footnote{For $10^6$ 
particles distributed uniformly inside a spherical region, 
the fraction of particles that lie near the boundary is $5\%$ and
for a a $3$ dimensional space and $79\%$ for a $6$ dimensional space.
}.

In FiEstAS the following correction is implemented: 
Suppose a leaf node having a particle at $\bm{x^p}$ 
has one of its surfaces in dimension $i$ either $x_i=x_{max}$ or 
$x_i=x_{min}$ as a boundary, then the boundary face is  
redefined such that its distance from the particle is same as
the distance of the other face from the particle.
For the former the redefinition is  $x_i= x_i^p + (x^{p}-x^{min})$ 
and for the later  $x_i=x^p_i-(x^{max}-x^p_i)$.
If both the faces lie on the boundary then the scheme fails to 
apply the correction. Moreover for small sub halos embedded in 
a bigger halo the sub halos have lower velocity dispersion and  
occupy a smaller region in velocity space, hence its boundary needs
to be corrected even though it is not directly derived from the 
global boundary. A similar situation also arises near the  
center of the halos for which the circular velocity $V_c(r) \rightarrow 0$ as
$r \rightarrow 0$. Moreover in EnBiD we need to calculate the entropy for each node and the
boundary effects might decrease the entropy of the system spuriously.
Consequently, a boundary correction needs to be applied to each node 
during the tessellation, and not just to the leaf node at the end 
of tessellation. 
In EnBiD for each node  having more than a given threshold $n_b$ of particles, 
a node is checked for boundary correction before the calculation of entropy.
In a given dimension if  $l_{max}$ and $l_{min}$ are the maximum and minimum 
co-ordinates of a node and $x_{max}$ and $x_{min}$, the corresponding 
 maximum and minimum co-ordinates of the particles inside it, then
 a boundary correction is applied if simultaneously
 \be
 (l_{max}-x_{max}) > f_b \frac{x_{max}-x_{min}}{n_{node}-1}
 \ee
and 
  \be
  (x_{min}-l_{min}) > f_b \frac{x_{max}-x_{min}}{n_{node}-1}
  \ee
where $f_b$ is a constant factor. This is effective in detecting 
embedded structures. To check for corrections applicable for 
only one face, the value of $f_b$ is chosen to be $5$ times higher.
For cubical cells in real and velocity space $f_b=0.5 N^{1/d}$ was found to
give optimal results, where $N$ is the total number of particles in the
system. For general decomposition the corresponding value of $f_b$ was
found to be $2.0N^{1/d}$.
 The node boundary $l_{max}$ and $l_{min}$ are corrected as
 \be
 l_{max} \rightarrow x_{max}+\frac{x_{max}-x_{min}}{n_{node}-1} \\
 l_{min} \rightarrow x_{min}-\frac{x_{max}-x_{min}}{n_{node}-1}
 \ee
 where $(x_{max}-x_{min})/(n_{node}-1)$ is the expected mean inter-particle
 separation.

The choice of $n_b$ is dictated by two factors  a)
if $d$ is the number of dimensions of the space
then a minimum of $d+1$ particles are needed to define a geometry 
in that space so  we set $n_b \ge d+1$.
If the number of particles in a node are too small 
this leads to Poisson errors in the calculation of the inter-particle
separation so we impose a lower limit of $n_b=7$.

\subsection{Smoothing}
The un-smoothed density estimates have a large dispersion which cannot 
be reduced even by increasing the number of particles. 
By smoothing this dispersion can be reduced provided the density does
not vary significantly over the smoothing region.
We test two different smoothing techniques. The FiEstAS smoothing
as proposed by \citep{asca05} and the kernel based scheme (KDE).

In FiEstAS smoothing, first the density of each node is calculated assuming that 
the mass of each particle is distributed uniformly over its leaf node. 
Next  the volume $V_s$ centered on that point which
encompasses a given smoothing mass $M_s$ is calculated. 
The density estimate is then
given by $\rho=M_s/V_s$. For Cubic tessellation the smoothing cells are 
also chosen to be exactly cubical in the real and velocity subspaces.
To calculate $V_s$ an iterative procedure is used. We start with a
hyper-box having boundaries in the $i$-th dimension at $x_i\pm\Delta_i$,
$\Delta_i$ being the distance to the closest hyper plane along $i$-th
axis of the leaf node containing the point $\bm{x}$. $\Delta_i$ is
then doubled until the  mass enclosed  by smoothing box $M<M_s$ and 
then the interval is halved repeatedly till $|(M-M_s)/M_s| \leq \eta_{tol}$
where $\eta_{tol}$ is a tolerance parameter.  
Our experiments show that a tolerance parameter of 0.1 gives satisfactory results.
Although in FiEstAS the smoothing mass $M_s=10 m_p$ is chosen, we find that
choosing $M_s=2 m_p$ gives a higher resolution, while 
not compromising much on the noise reduction. 

In Kernel smoothing a fixed number of nearest neighbors around the 
point of interest are identified and the density is computed 
by summing over the contributions of each of the neighbors by
using a kernel function. This is known as the adaptive kernel 
smoothing since the smoothing length is $\propto \rho^{1/d}$, 
$\rho$ being the density in a $d$ dimensional space.
The kernel function can be spherical 
of the form of $W(u$), $u = \sqrt{\sum_{i=1}^{d} u_i^2 }$ being the distance of the neighbor from 
the center and $u_i$ the corresponding co-ordinates in a $d$ dimensional
space, or of the form of $\Pi_{i=1}^{d} W(u_i)$ known as the product kernel.
The standard kernel scheme provides a much poorer 
estimate of the phase space density, since a global metric is 
usually unsuitable in accounting 
for the complex real and velocity structure encountered in many
astrophysical systems. 
However, with a method like EnBiD we can
determine the appropriate metric at each point in space and thus
force the co-variance to be approximately same along all dimensions. 
At any given point the correct metric can be calculated by 
determining the sides of the leaf node which encompasses that point,
followed by a coordinate transformation such that the node is
transformed into a cube. As we illustrate in the appendix, the 
kernel density estimator can have a  significant bias in the estimated densities.
The results we show here are after correcting for this bias. 
We tested and compared the use of spline and the 
Epanechnikov kernel function and found the later to be more efficient.
For all our analysis we use the Epanechnikov kernel function.
Bias correction and other details pertaining to kernel based methods 
e.g the number of smoothing neighbors are given in the Appendix.
The algorithm implemented in EnBiD for nearest neighbor search 
is based on the algorithm of SMOOTH \citep{Sta95}.

Although the length of the sides of a node provides an accurate 
estimate of the metric but when trying to smooth over a region,  
the smoothing region might exceed the boundaries of the actual particle 
distribution. The smoothing lengths in such case needs to be appropriately 
redefined. This situation arises in cases where a  dimension has very less
entropy and has been split many times or near the boundaries of the system 
where the metric has not been accurately determined.
In a given dimension let  $l_{max}$ and $l_{min}$ be the maximum and minimum 
co-ordinates of a smoothing box or a sphere encompassing a fixed number of
neighbors $N_{ngb}$ and $x_{max}$ and $x_{min}$ the maximum and minimum
co-ordinates of the particles inside it. 
A smoothing length correction is applied to the box, if simultaneously 
the distance to both the right and left boundaries
given by 
\be
(l_{max}-x_{max}) > 25(x_{max}-x_{min})/N_{ngb} \\
(x_{min}-l_{min}) > 25(x_{max}-x_{min})/N_{ngb}
\ee
where $N_{ngb}$ is the number of smoothing
neighbors. The metric is redefined with $l_{max}$ and $l_{min}$ 
set to $x_{max}$ and $x_{min}$. 
For FiEstAS smoothing also we implement a similar smoothing volume correction. 
For a given smoothing box of volume $V_s$, if $m_i$ is the mass contributed 
by a leaf node to the smoothing box and $v_i$ its corresponding volume 
that falls within the box, then instead of calculating  the density 
as $\rho=\sum m_i / V_s$, we calculate it as $\rho=(\sum m_i )/(\sum v_i)$.
This correction is only applied if $(\sum v_i)/V_s < 0.5$.

\section{Tests}
To test the accuracy of the results we generate test data with a given
density distribution in a $d$ dimensional space and then perform a
comparison with the density estimates given by the code.
We employ systems which have an analytical expression of $6$
dimensional phase space density $f$, namely an isotropic Hernquist
sphere (c.f. \citet{asca05}) and an isotropic halo with a Maxwellian
velocity distribution (c.f. \citet{arad04} ).
The test cases are  generated by discrete random sampling of this density
function $f$ using a fixed number of particles $N$.
We show here results of tests done in 6 dimensions only and with
boundary correction and smoothing. Results pertaining to $3$
dimensions and effects of boundary correction and smoothing are
discussed in detail in \citet{asca05}.

\begin{figure}
    \centering \includegraphics[width=0.45\textwidth]{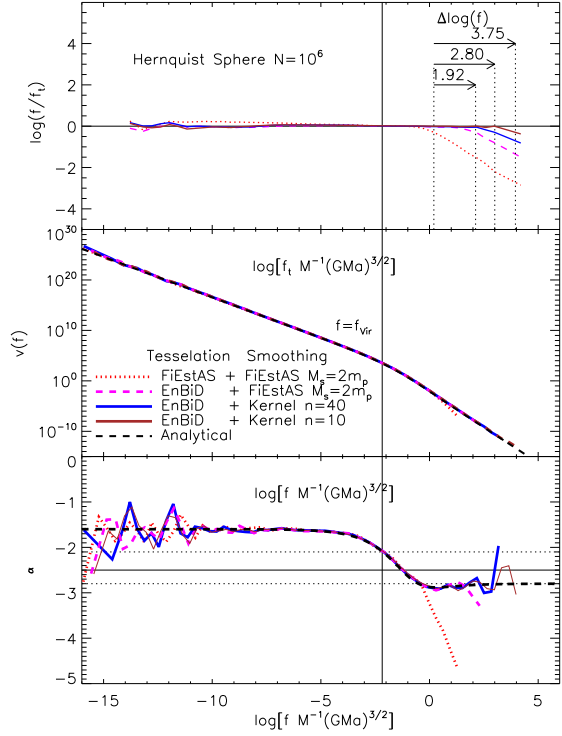}
  \caption{Dependence of fraction $f/f_t$ on $f_t$ and $v(f)$ and $\alpha(f)$
    on $f$ for a Hernquist
    sphere with $N=10^6$ particles obtained by different algorithms
    for density estimation. Vertical dotted lines mark the position
    where $f/f_t=0.5$. EnBiD resolves the high-density regions 
    better by about $2$ decades in density. Kernel Smoothing using the metric
    as determined by EnBiD performs even better (a gain in resolution of about
    3-4 decades). Using a smaller number of smoothing neighbors results in 
    higher resolution.
\label{fig:fft_h}}
\end{figure}

\begin{figure}
    \centering \includegraphics[width=0.45\textwidth]{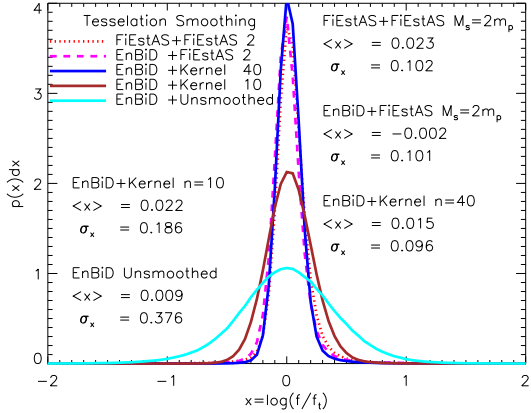}
  \caption[Probability distribution $P(\log(f/f_t))$ for a
    Hernquist sphere]{ Probability distribution$P(\log(f/f_t))$ for a
    Hernquist sphere with $N=10^6$ particles obtained by different algorithms
    for density estimation. 
\label{fig:pff_h}}
\end{figure}

\subsection{Hernquist Sphere} \label{sec:hern}
For a \citet{1990ApJ...356..359H} sphere of total mass $M$ and scale length $a$ the 
real space density is given by
\ben
\rho(r) & = & \frac{M/(2\pi a^3)}{(r/a)(1+r/a)^3} 
\label{eq:rho_he}
\een
and gravitational potential is given by     
\ben
\phi(r)=-\frac{GM}{a}\frac{1}{1+r/a}.
\een

The phase space density as a function of energy $E=v^2/2+\phi(r)$ is
\be
f(E)&=&\frac{M/a^3}{4\pi^3 (2GM/a)^{3/2}} \times \\
&  &
\frac{3\sin^{-1}q+q\sqrt{1-q^2}(1-2q^2)(8q^4-8q^2-3)}{(1-q^2)^{5/2}} \
\ \ \ \ \ \ \
\label{eq:hernf}
\ee
where
\ben
q = \sqrt{-\frac{E}{GM/a}}.
\een

First we generate a random realization in real space corresponding to 
density given by \equ{rho_he}. Then we use von Neumann rejection
technique to generate the velocities that sample the distribution \citet{1992nrca.book.....P}.
\ben
p(v)dv=\frac{4\pi}{\rho(r)}f(v^2/2+\Phi(r))v^2dv
\een
Further details can be found in \citet{asca05}.
For calculating the virial quantities of a Hernquist sphere
we use $c=R_{Vir}/a=4.0$ which roughly corresponds to an $NFW$ halo
with $c=8.0$.

In top panel of \fig{fft_h}  we plot the ratio of numerically estimated phase density  $f$
evaluated by the respective method to the analytical phase space density $f_t$ 
, as a function of $f_t$ for a Hernquist sphere
sampled with $10^6$ particles. 
$f$ is calculated by binning the particles in $80$ logarithmically spaced 
bins in $f_t$ with at least $5$ particles per bin and then evaluating the 
mean value of the estimated density of all the particles in the bin.

Ideally one expects the plot to be a
straight line with $f/f_t=1$. It can be seen from the figure that the
density is well reproduced for most of the halo for about $18$ decades
in density except near the very center where the density is very
high. Both FiEstAS and EnBiD tessellation, followed by FiEstAS 
smoothing with $M_s=2m_p$, underestimate the density in the region of very 
high density, however when compared to FiEstAS tessellation the
high density cusp is resolved better by EnBiD by about 2 decades in density.
In real space there is more variation of density 
as compared to velocity space. EnBiD accounts for this by
allocating more divisions in real space thereby achieving higher 
spatial resolution, whereas FiEstAS gives equal weight 
to both spaces and ends up thus compromising the spatial resolution.
When kernel smoothing is employed along with metric as determined by EnBiD
tessellation (EnBiD+Kernel Smooth), there is a further gain in resolution by
about $3$ and $4$ decades for smoothing neighbors $n=40$ and $n=10$
respectively. Lowering the 
number of smoothing neighbors results in higher resolution.

Next we compare the volume distribution function $v(f)$ as reproduced
by the code. Numerically $v(f)$ is evaluated by binning the particles as
before in logarithmically spaced bins of $f$.
If $m_{bin}$ is the mass of all the particles 
in the $i$ th bin, the density  of the bin being $f_{bin}=(f_{i+1}+f_i)/2$   
then $v(f_{bin})=(m_{bin}/f_{bin})/(f_{i+1}-f_i)$. 
Statistical error in each bin 
is given by $\Delta f =f_{bin}-<f_{bin}>$ (where $<f_{bin}>$ is the mean 
value of density of all the particles in the bin).
Analytically the volume distribution function is given by
\be
v[f(E)]=\frac{g(E)}{f'(E)}
\label{eq:hernv}
\ee
where $g(E)$ is the density of states. For a Hernquist sphere  
\ben
g(E) &  = &  \frac{2\pi^2a^3(2GM/a)^{1/2}}{3q^5}  [3(8q^4-4q^2+1)cos^{-1}q \\
&  & -q(1-q^2)^{1/2}(4q^2-1)(2q^2+3)] 
\een

It can be seen from middle panel of \fig{fft_h} that $v(f)$ is well reproduced by
both FiEstAS and EnBiD. However, in the high density region FiEstAS 
underestimates $v(f)$ which results in steepening of the volume
distribution function at the high $f$ end, while EnBiD estimates the 
$v(f)$ accurately to much higher densities.

This can be seen more clearly in lower panel of \fig{fft_h} where we plot the
logarithmic slope denoted by $\alpha$ of the volume distribution
function as function of density $f$. 
\ben
\alpha=\frac{d\log(v(f))}{d\log(f)}
\een
FiEstAS can reproduce the slope parameter 
$\alpha$ only till $f/f_{Vir}=10^2$ whereas EnBiD can reproduce it
till $f/f_{Vir}=10^4$ and EnBiD+Kernel Smooth can reproduce it 
till  $f/f_{Vir}=10^5$ and $f/f_{Vir}=10^6$, for smoothing neighbors
$n=40$ and $10$ respectively.

In order to get an estimate of the dispersion in the reproduced values of 
$f$ and in order to check the effectiveness of smoothing we plot in
\fig{pff_h} the probability distribution of $f/f_t$. The distribution
can be fitted with a log-normal distribution and the fit parameters 
are also shown in the figure. The bias is less than $0.03$ dex for all
the methods.The un-smoothed estimates have a  dispersion of $0.37$ dex.
FiEstAS smoothing with $M_s=2$ is equivalent to Kernel smoothing with
smoothing neighbors $n=40$. Both of them have a dispersion of about $0.1$ dex.
For kernel smoothing lowering the smoothing neighbors to $n=10$ results in 
an increase in dispersion to $0.18$ dex.

\begin{figure}
    \centering \includegraphics[width=0.45\textwidth]{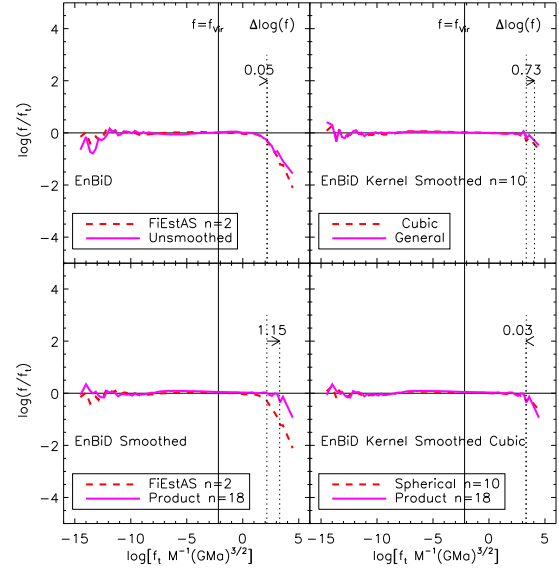}
  \caption{Dependence of fraction $f/f_t$ on $f_t$ for a Hernquist
    sphere with $N=10^6$ particles obtained by different algorithms
    for density estimation. 
\label{fig:fft2_h}}
\end{figure}

The EnBiD tessellation in the results as analyzed above was done with {\it Cubic 
Cells} in real and velocity space. In top right panel of \fig{fft2_h} we
compare the results as obtained 
with {\it General} decomposition where each dimension is treated independently.
Kernel smoothing with smoothing neighbors $n=10$ was employed for both of them.
The estimates are nearly identical. There is a slight gain in resolution
but the estimates with {\it General} decomposition were also found to have 
a slightly higher dispersion in the estimates.
In bottom right panel we compare the result of smoothing between a product 
kernel and a spherical kernel. There is very little difference between the estimates.
The number of neighbors were chosen so as to have identical dispersions 
in both the estimates. When using the kernel in product form about 
double the number of neighbors are needed to obtain identical dispersion.

In top left panel we compare the un-smoothed densities with 
FiEstAS smoothed densities. For both of them EnBiD scheme is used for 
tessellation. The un-smoothed estimates are the densities 
as determined from the volume of the leaf nodes generated by the 
tessellation procedure. The FiEstAS smoothing only reduces the 
dispersion the resolution remains nearly unaltered. The resolution and 
accuracy is essentially determined by the density of the leaf nodes.
Next we compare the FiEstAS smoothing with cloud in cell scheme  \citep{HE} 
of density estimation. 
The cloud in cell (CIC) method of density estimation is a special 
case of smoothing with a product kernel along with a linear kernel
function $W(u) \propto (1-u)$.
Although the FiEstAS smoothing is similar to the cloud in a cell 
scheme of density estimation but is still unique in its own respect. 
The main difference being 
that the clouds which are the leaf nodes in case of FiEstAS smoothing 
are disjoint whereas in cloud in cell  scheme or in general for Kernel 
based schemes they are overlapping. They can smooth over much smaller regions 
and 
hence achieve higher resolution as compared to FiEstAS smoothing.
 In bottom left panel we plot the estimates of FiEstAS smoothing 
alongside the estimates as obtained with product kernel  
with smoothing neighbors n=18. Instead of a linear kernel function 
we use the Epanechnikov kernel. 
It can be seen from the figure that the resolution achieved with 
product kernel is higher as compared to that of FiEstAS smoothing.

When decomposition was done alternately in each dimension the median 
criterion gave more accurate results. However for EnBiD decomposition 
choosing the splitting point at either the  
mean or the median both gave similar results for density estimation 
of a hernquist sphere, but for a system having substructures the mean 
criterion gave better results.  
For all our analysis unless otherwise mentioned, for evaluating
phase space densities we use EnBiD decomposition with {\it Cubic Cells} to 
determine the metric and then use the method of spherical kernel 
smoothing for calculating densities. The mean criterion is used for 
choosing the splitting point. The number of smoothing neighbors $n$ is
chosen to be $40$, although  choosing $n=10$ gives higher resolution 
but it also has higher dispersion which means that the volume 
distribution function will be smoothed out below the scale 
set by the dispersion (see \se{dekel} for more explanation).

\begin{table}
  \caption[Comparison of time needed to calculate densities by
    various methods]{Comparison of time needed to calculate densities by
    various methods: This is the time taken to calculate the six
  dimensional phase space density of a Hernquist sphere with $10^6$
  particles on an AMD XP2000+ processor having a clock speed of 1666.67 MHz}
\begin{center}
  \begin{tabular}{@{}llccc}
\hline
\hline
  Method   &    &   Tree   &  Smooth &  Total \\
Tessellation   & Smoothing &  Building   &   &   \\
\hline
AD Delaunay & & &
& $1week$ \\
AB FiEstAS &FiEstAS $M_s=10m_p$& $4$s & $730$s & $724$s \\ 
FiEstAS &FiEstAS $M_s=10m_p$& $8$s & $522$s & $532$s \\ 
FiEstAS &FiEstAS $M_s=2m_p$& $8$s & $306$s & $317$s \\ 
EnBiD&FiEstAS $M_s=2m_p$& $19$s & $336$s & $356$s \\ 
EnBiD&Kernel $N_{sm}=40$& $19$s & $843$s & $863$s \\
EnBiD&Kernel $N_{sm}=10$& $19$s & $405$s & $426$s \\
\hline
\end{tabular}
\end{center}
\label{tb:times}
\end{table}

In \tab{times} we compare the CPU time needed to estimate the phase space 
density of $10^6$ particles in a Hernquist sphere by various methods
and techniques. The time as reported  by \citet{asca05} for FiEstAS 
is labeled as AB FiEstAS and the time as reported by \citet{arad04} for Delaunay 
Tessellation method as AD Delaunay.
It can be seen that most of the time is needed for
smoothing. For both FiEstAS and  Kernel smoothing, increasing 
the smoothing mass 
or the number of smoothing neighbors, increases the time. 
Our implementation of FiEstAS smoothing is slightly faster 
as compared to that of \citet{asca05} due to better cache utilization. 
This is achieved by ordering the  particles 
just as they are arranged in the binary tree. 
The kernel smoothing which gives more accurate results requires a modest 
$20\%$ more time as compared to the time reported in \citet{asca05} for
FiEstAS. For median splitting it is possible to speed up the neighbor 
search by about $10\%$.

\begin{figure}
    \centering \includegraphics[width=0.45\textwidth]{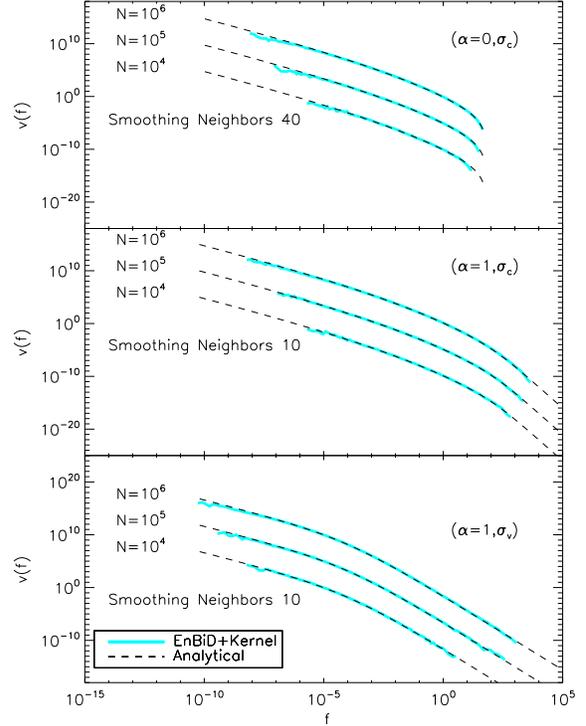}
  \caption[The cumulative distribution of $f/f_t$ as measured in
    different bins of $f_t$ for three different mock systems.]{ The cumulative distribution of $f/f_t$ as measured in
    different bins of $f_t$ for three different mock systems. The
    density is progressively overestimated in low density regions. 
\label{fig:vff_d}}
\end{figure}

\subsection{Maxwellian Velocity Distribution Models} \label{sec:dekel}
For these models the phase space density is given by 
\begin{eqnarray}
f(r,v) & = & \rho(r)[2\pi\sigma(r)^2]^{-3/2}e^{v^2/2\sigma(r)^2}
\end{eqnarray} 
where $\rho(r)$ is the real space density given by
\begin{eqnarray*}
\rho(r) & = & \frac{e^{-r/(5r_s)}}{(r/r_s)^{\alpha}(1+r/r_s)^{3-\alpha}}
\end{eqnarray*} 
The  velocity dispersion is assumed to be either constant with 
$\sigma_c(r)=0.1$  or variable with $\sigma_v(r)=\sqrt{M(r)/r}$. 
We generate models with $\alpha=0$ and $\alpha=1$.

The volume distribution function $v(f)$ for such systems is given by
\begin{eqnarray}
v(f) & = & \frac{(4\pi)^2}{f}\int_{0}^{r(f)} r^2 \sigma(r)^3 \sqrt{2\log\frac{f(r)}{f}}dr
\end{eqnarray} 
where 
\begin{eqnarray}
f(r)=\frac{\rho(r)}{[2\pi\sigma(r)^2]^{3/2}}
\end{eqnarray} 

In \fig{vff_d} we show the volume distribution function as recovered
by EnBiD along with kernel smoothing for three different models 1) $\alpha=0,\sigma_c$  2)
$\alpha=1,\sigma_c$ and 3) $\alpha=1,\sigma_v$ and with three
different particle resolutions $N=10^4$ ,  $N=10^5$ and $N=10^6$. 
For the highest resolution the volume distribution can be
recovered for about $9$ to $13$ decades in $f$. The range of 
densities over which the $v(f)$ is reliably recovered increases with
increasing particle number. For systems with a sharp transition in
slope of $v(f)$ for example $\alpha=0,\sigma_c$ system, Delaunay
Tessellation was found to significantly over-estimate $v(f)$ (Fig-A2 \citet{arad04}),
because the measured $v(f)$ can be thought of as a convolution of the
exact $v_t(f)$ with a fixed window function $p(f/f_t)$. The narrower the 
$p(f/f_t)$ the closer is $v(f)$ to $v_t(f)$. If  $v_t(f)$ varies
significantly over scales smaller than the width of $p(f/f_t)$ the shape
of recovered $v(f)$ will be affected. The $v(f)$ will be over-estimated 
for a system with a sharp change in the slope of $v(f)$. Moreover
due to the width of $p(f/f_t)$ the effective cut-off value of $f$ is also
higher as compared to the theoretically expected upper bound. A
bias in $p(f/f_t)$ will also affect the results.
Delaunay Tessellation estimates have a width of about one decade in the
distribution of $p(f/f_t)$.
With EnBiD (using smoothing neighbors $n=40$) 
for  $\alpha=0,\sigma_c$ system at the high $f$ end there is very 
little width in the recovered values of $f$, this is the reason that $v(f)$ is
recovered better by EnBiD as compared to Delaunay Tessellation (\fig{vff_d}).
For other systems the range of $f$ over which $v(f)$ is recovered is
slightly higher for EnBiD (using smoothing neighbors $n=10$) as compared to
Delaunay Tessellation (Fig-A2 and A3 \citet{arad04}).

\begin{figure}
  \centering \includegraphics[width=0.45\textwidth]{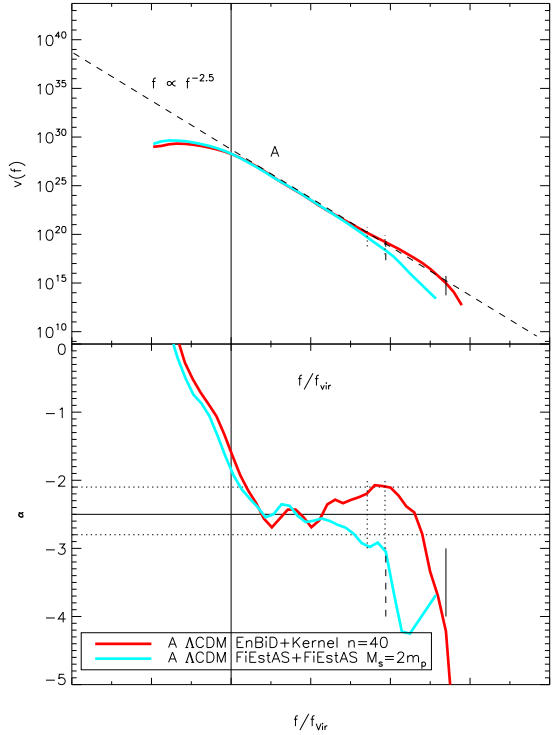}
  \caption[Comparison between density estimators EnBiD and FiEstAS in
    extracting the
    volume distribution function and the slope parameter $\alpha$ of a
    $\Lambda$CDM halo.]{ Comparison between density estimators EnBiD and FiEstAS in
    extracting the
    volume distribution function and the slope parameter $\alpha$ of a
    $\Lambda$CDM halo. The vertical dotted lines mark the position of
    $f=f_{discr}$  and $f=f_{relax}$ ($f_{discr}$ being less than
    $f_{relax}$).     The solid vertical line marks
    the point where statistical errors ( $\Delta f/f >0.1$ in a bin)
    in calculation of    $v(f)$ become  important (dashed one for
    FiEstAS) .  The estimator FiEstAS fails to resolve the high
    density regions accurately and    this results in steepening of
    the $v(f)$ profile.
\label{fig:fien_r}}
\end{figure}

\section{Phase Space Structure of Dark Matter Halos}
We are now applying our tools to the phase space structure of 
virialized dark matter halos in a concordance $\Lambda$CDM universe
\citep[]{2003ApJS..148..175S,2003ApJ...586L...1M}. 
The structure of these halos in real
space has been studied in great detail over the past decade and the
radial density profile is known to follow an almost universal form
known as the NFW \citep{1996ApJ...462..563N,1997ApJ...490..493N}
profile (see however \citet{2004MNRAS.349.1039N} for a new $\alpha$ profile).
\be
\rho(r)=\frac{\rho_s}{(r/r_s)(1+r/r_s)^2}
\ee
The dark matter particles are
collisionless and obey the collisionless Boltzmann equations.
For a collisionless spherical system in equilibrium  with a given
density profile $\rho(r)$ the phase space density $f(r,v)$ can be
calculated using the Eddington equation \citep{Binney87}.

\begin{eqnarray*}
f(\varepsilon) & =& \frac{1}{\sqrt{8}\pi^2}\left[ \int_{0}^{\varepsilon}
  \frac{d^2\rho}{d\psi^2} \frac{d\psi}{\sqrt{\varepsilon-\psi}}-
  \frac{1}{\varepsilon}\left(\frac{d\rho}{d\psi}\right)_{\psi=0} \right]
\end{eqnarray*}

Since $f$ is a function of six variables it is hard to study except in
cases where there are isolated integrals of motion which reduce
the number of independent variables. To study the structure of phase
space density, the function $v(f)$ is introduced which is the volume distribution
function of $f$. $v(f)df$ is the volume of phase space
occupied by phase space elements having density between $f$ to $f+df$.
\citet{arad04} calculated the phase space density using Delaunay
Tessellation in 6 dimensions and studied the volume distribution function
of halos obtained from simulations. They found that $v(f)$ follows an
almost universal form which is a power law with slope $-2.5 \pm 0.05$
which is valid for about four decades from $f_{Vir}$ to $f_{Vir} 10^{4}$.
$f_{Vir}$ is an estimate of the phase space density in the outer parts of the halo.
\ben
f_{Vir} & = & \frac{\bar{\rho_{Vir}}}{\pi^{3/2} V_{Vir}^3} \\
& = & \left[\frac{3\Delta    \rho_c}{4\pi^4G^3}\right]^{1/2}\frac{1}{M_{Vir}} \\
& = &\frac{1.64\times10^{9}
  h^2\Msun\kpc^{-3}(\kms)^{-3}}{(M_{Vir}/\Msun h^{-1})} \ \ \ \ \ \textrm{ Using $\Delta=101$}
\een

This behavior was also found to be independent of redshift and the
mass of the halo. \citet{asca05} used the FiEstAS algorithm to
calculate the phase space densities and confirmed the above result and
in addition found slight deviations both at low and high $f$ end. 
At the low $f$ end (near $f_{Vir}$) the slope was found to be
flatter than $-2.5$ and at the high $f$ end it was found to be
significantly steeper.
At the high $f$ end there are two relevant numerical phase space 
densities, above which two-body relaxation and discreteness effects 
in simulations start dominating.
The phase space density above which the two body relaxation is shorter
than the age of the universe is given by \citep{2004MNRAS.348..977D}
\be
f_{relax} & = & \frac{0.34}{(2\pi)^{3/2}G^2\ln\Lambda}\frac{1}{m_p t_0} \\
& = &  \frac{1.94 \times 10^{7} h^2 \Msun \kpc^{-3} (\kms)^{-3}}
{(m_p/\Msun h^{-1})} 
\ee
The above value is obtained by assuming a Coulomb logarithm of $\ln\Lambda=6$
and using $t_0=14.5 Gyr$ as the age of the Universe.
The phase space density, above which the discreteness effects discussed
by \citet{2004MNRAS.350..939B} become important, is
\be
f_{discr} & = & \frac{(\Omega_m \rho_c)^2}{H_0^3 m_p} \\
& = & \frac{6.93 \times 10^{6} h^2 \Msun \kpc^{-3} (\kms)^{-3}} {(m_p/\Msun  h^{-1})}
\ee
Since the steepening was found to roughly coincide with 
these densities, this effect was attributed by \cite{asca05} 
to the numerical effects of the simulations.

We analyze $5$ halos at $z=0$ simulated in a  $\Lambda$CDM cosmology 
with  $\Omega_{\lambda}=0.7$, $\Omega_{m}=0.3$. 
To evaluate the phase space densities we use the EnBiD scheme 
along with kernel smoothing employing $n=40$ neighbors.
Halos A, B, and C were
isolated from a cosmological simulation of $128^3$
dark matter particles in a $32.5 h^{-1}\Mpc$ cube performed by $AP^3M$
code \citep{1991ApJ...368L..23C} and were then re-simulated at higher resolution from $z=50$ to $z=0$
using the code GADGET \citep{Springel01}.
Halo A' is a warm dark matter (WDM) realization of halo A
which was generated by suppressing power on scales smaller than the
size of the halo. Halo D is from a simulation done with an ART code
\citep{1997ApJS..111...73K} with a box size of $80 h^{-1}\Mpc$. Further
details are given in \tab{prop}. For calculating phase space densities 
we use the EnBiD tessellation scheme and smoothing is done with 
a spherical kernel employing $n=40$ neighbors.

\begin{table*}
  \caption[Properties of halos  whose phase
    space structure is analyzed]{Properties of halos whose phase
    space structure is analyzed here: $N_{cut}$ is the
    number of particles that lie within  a cutoff radius $R_{cut}$. 
    These are the particles that are used for calculating the volume
    distribution function $v(f)$ of the halo.
    }
\begin{center}
  \begin{tabular}{@{}ccccccccc}
\hline
\hline
Halo & $N_{\mathrm{cut}}$ & $R_{\mathrm{cut}}$ &  $R_{Vir}$ & $M_{Vir}$ & Hubble $h$ & Softening &   Code &  Power  \\
     &                  &                  &           &          & 
       Parameter  &           &      &  Spectrum  \\          
     &                  &  $\kpc$          &  $\kpc$   &  $\Msun$ &            & $\kpc$    &      &           \\
\hline
$A$ & $6.2 \times 10^5$ & $348.9$  & $348.9$  & $2.11 \times 10^{12}$ & $0.65$ & $0.30$ &  GADGET &  $\Lambda$CDM \\
$B$ & $6.1 \times 10^5$ & $692.6$  & $692.6$  & $1.65 \times 10^{13}$ & $0.65$ & $1.53$ &  GADGET &  $\Lambda$CDM \\
$C$ & $3.2 \times 10^5$ & $463.2$  & $463.2$  & $4.93 \times 10^{12}$ & $0.65$ & $1.53$ &  GADGET  &  $\Lambda$CDM \\
$D$ & $6.5 \times 10^6$ & $1854.0$  & $1854.0$  & $ 3.2 \times 10^{14}$ & $0.70$ &     & ART    &  $\Lambda$CDM  \\
$A'$ & $4.5 \times 10^5$ & $312.1$  & $312.1$  & $1.51  \times 10^{12}$ & $0.65$ & $0.30$ &  GADGET &  $WDM$ \\
\hline
\end{tabular}
\end{center}
\label{tb:prop}
\end{table*}

\begin{figure}
    \centering \includegraphics[width=0.45\textwidth]{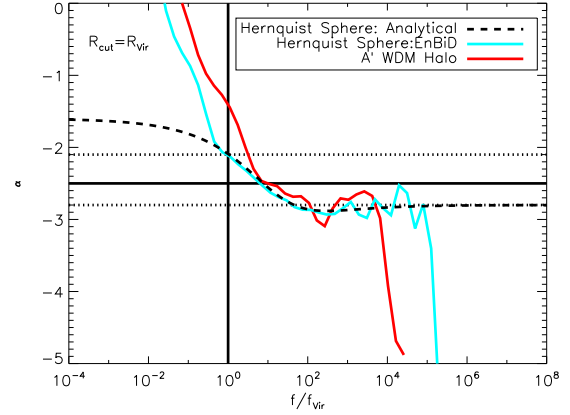}
  \caption[Effect of truncation on the slope parameter $\alpha$
    as extracted from a mock Hernquist sphere.]{ Effect of truncation on the slope parameter $\alpha$
    as extracted from a mock Hernquist sphere. The dotted line is
    the true analytical profile of a Hernquist sphere. 
    For a halo whose $R_{cut}=R_{Vir}$
    and $c=4.0$ the $\alpha$ profile can only be extracted till
    $f=f_{Vir}$. The rise in value of $\alpha$ beyond this is
    due to truncation of the halo. 
    The thin dark line is for a WDM halo 
    obtained from simulations. 
\label{fig:alf_ht}}
\end{figure}

\begin{figure}
    \centering \includegraphics[width=0.45\textwidth]{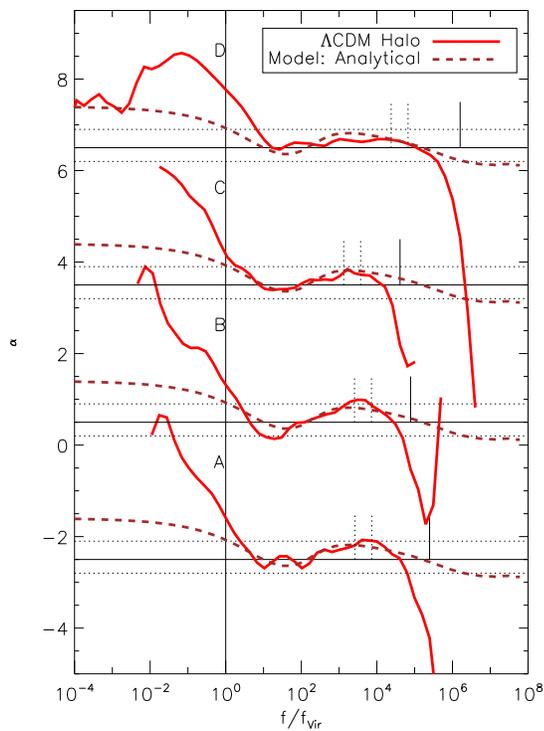}
  \caption[The dependence of slope parameter $\alpha$ on $f$ 
    for four halos obtained from $\Lambda$CDM simulations.]{The dependence of slope parameter $\alpha$ on $f$ 
    for four halos obtained from $\Lambda$CDM simulations. 
    The values of $\alpha$ for halos B,C and D have been shifted by 3,6 and 9
    respectively for the sake of clarity.
An explanation of
    vertical lines is given in  \fig{fien_r}. The dashed line represents the
    analytical profile of the parent $+$ substructure model. The
    dotted line is the profile as estimated by EnBiD for the synthetic
    realization of the corresponding model. The parameter
    $\alpha$ does not have a constant value of $-2.5$ but has a dip
    and rise and is bounded between $-2.8$ (the asymptotic value of a
    Hernquist sphere) and $-2.1$ (the value predicted by the AD Toy model)
    which are indicated by horizontal dotted lines. 
    \label{fig:alf_r}}
\end{figure}

\begin{figure}
    \centering \includegraphics[width=0.45\textwidth]{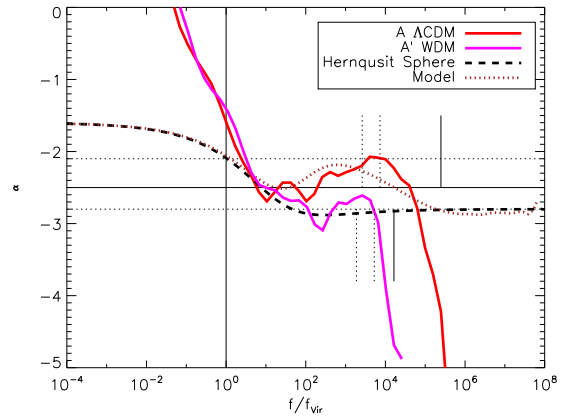}
  \caption[The dependence of slope parameter $\alpha$ on $f$
    for a $\Lambda$CDM and a WDM halo.]{ The dependence of slope parameter $\alpha$ on $f$
    for a $\Lambda$CDM and a WDM halo. The behavior of WDM halo profile is in agreement
    with that of a Hernquist Sphere while that of $\Lambda$CDM halo is close
    to that of a parent $+$ substructure model. The vertical lines mark the position
    of $f_{stat},f_{relax}$ and $f_{discr}$ for the $\Lambda$CDM halo.
\label{fig:alf_cw}}
\end{figure}

It can be seen from \fig{fien_r} that at the high $f$ end 
there are differences between the phase space properties of halos as 
reproduced by EnBiD (kernel smoothing using $40$ neighbors) and FiEstAS
(FiEstAS smoothing using smoothing mass $M_s=2m_p$). We argue that the 
steepening of the volume
distribution function as found by \citet{asca05} is probably an
artifact of the FiEstAS algorithm since such a 
steepening also appears in tests done with a pure Hernquist sphere
(\se{hern}). 
For EnBiD we do not see such steepening; on the contrary, we see a slight hump.
This however does not preclude the association of discreteness and 
relaxation effects with the phase space structure of halos. 
Since we do not know the real phase
space density of the halo it is difficult to disentangle any such
effect from the effect of the estimator. 
For a WDM halo whose profile we expect to be the same as that of a
Hernquist sphere we do see a sudden change in slope at around
$f_{relax}$ (\fig{alf_cw}). 
Also the slope parameter of $\Lambda$CDM halos 
have a maximum which is around $f_{discr}$ and beyond this
it starts to fall off  \fig{alf_r}.
At the
low $f$ end the flatness in $v(f)$ profile is partly due to the 
truncation of the halo at a finite radius $r=R_{Vir}$. This is 
demonstrated in \fig{alf_ht}
where for a synthetic Hernquist sphere with $R_{cut}=R_{Vir}$ 
the $v(f)$ profile is found to flatten out beyond $f=f_{Vir}$ and $\alpha(f)$
rises sharply. 
The cosmological  halos exhibit a flattening that is more
pronounced than the synthetic halos. This suggests that their
structure is slightly different from that of an equilibrium spherical
model corresponding to a given density profile. Models with
anisotropy in the velocity dispersion also do not seem to suggest any 
extra flattening of the $v(f)$  profile.  
One possibility which was suggested earlier \citep{asca05} was that this
could be due to depletion of low density phase space by the presence 
of high density sub halos co-occupying the same space. This can be
ruled out as the low $f$ behavior of a WDM halo that does not exhibit 
significant substructure is identical to that of a $\Lambda$CDM halo.

\begin{figure}
    \centering \includegraphics[width=0.45\textwidth]{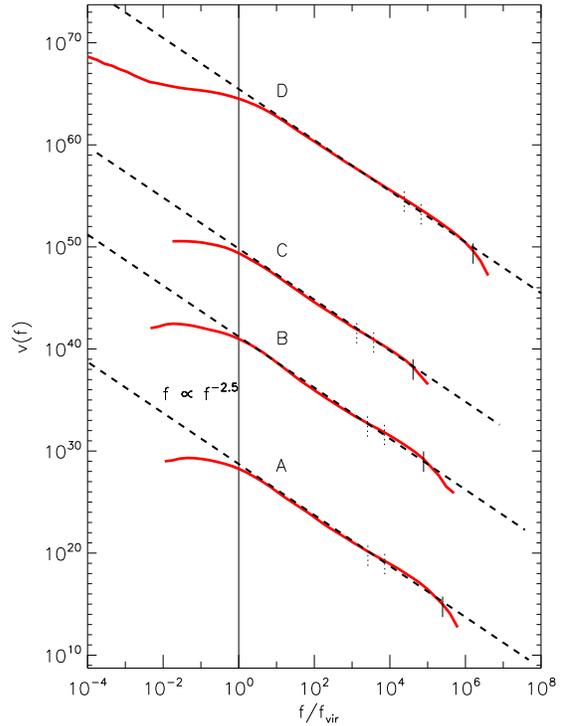}
  \caption[Volume distribution function of phase space density, 
    $v(f)$ for four halos obtained from $\Lambda$CDM simulations.]{ Volume distribution function of phase space density, 
    $v(f)$ for four halos obtained from $\Lambda$CDM simulations. 
    The values of $v(f)$ for halos B,C and D have been shifted by 10,20 and 30
    decades respectively for the sake of clarity. For reference
    $v(f) \propto f^{-2.5}$ curve (matched at $f/f_{Vir}=10$ )is plotted by a dotted line.
    An explanation of vertical lines is given in
    \fig{fien_r}.
\label{fig:vff_r}}
\end{figure}

Next we analyze the phase space structure of halos simulated in a
$\Lambda$CDM cosmology. We see the existence of a slight departure
from the constant power law behavior at the high $f$ end (\fig{vff_r}). 
The slope parameter $\alpha$ (\fig{alf_r}) has a minimum at around
$f/f_{Vir}=10$ and then it rises reaching a peak at around
$f/f_{Vir}=10^4$. Beyond this it starts to falls off.

\begin{figure}
    \centering \includegraphics[width=0.45\textwidth]{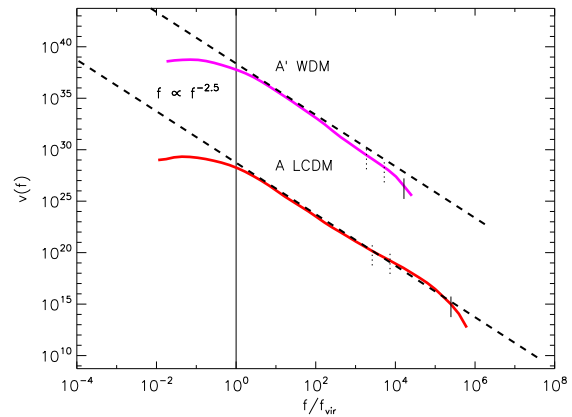}
  \caption[The volume distribution function of phase space density 
    $v(f)$ for a $\Lambda$CDM and a WDM halo.]{ The volume distribution function of phase space density 
    $v(f)$ for a $\Lambda$CDM and a WDM halo. The WDM profile has been shifted
    vertically  by 10 decades.
    An explanation of    vertical lines is given in  \fig{fien_r}.
    The WDM profile is significantly steeper in high density regions
    as compared to $v(f) \propto f^{-2.5}$ behavior which is
    indicated by a dashed line.
\label{fig:vff_cw}}
\end{figure}

In order to check whether the power law type behavior of the volume 
distribution function is due to the substructure or whether  it is 
associated with the virialization process we simulated a WDM halo whose 
power on small
scales has been suppressed and we find that it has a steeper slope at
the high $f$ end (\fig{vff_cw}). Its slope parameter $\alpha$ as a
function of $f$ is roughly consistent with that of a Hernquist sphere 
(\fig{alf_cw}). 
This suggests that the shape of the volume distribution function
is governed by the amount of substructure and its mass function.

\begin{figure}
  \centering \includegraphics[width=0.45\textwidth]{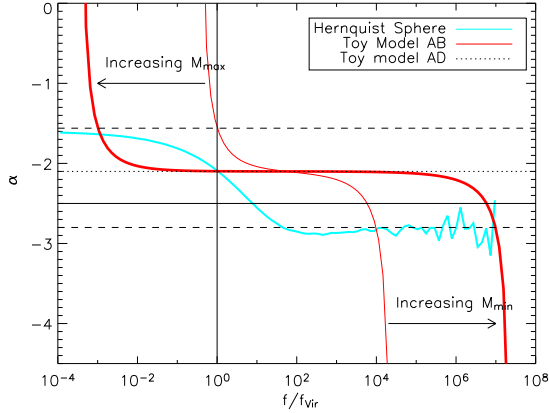}
  \caption[  Dependence of slope parameter $\alpha$ on $f$ as
    predicted by the toy model]{  Dependence of slope parameter $\alpha$ on $f$ as
    predicted by the toy model proposed in \citet{arad04} (AD) and the
    subsequent modification suggested by \citet{asca05} (AB). In
    the limit the parameter $m_{min} \rightarrow 0$ and parameter
    $m_{max} \rightarrow \infty$ the AB model goes over to AD model.
\label{fig:alf1}}
\end{figure}

\subsection{A Toy Model:Superposition of Sub Halos}
An elegant toy model to explain the near
power law behavior of the volume distribution function of simulated 
$\Lambda$CDM halos  was proposed by \citet{arad04} (model AD) 
In this model the halo is assumed to be made up of a superposition of
sub halos with a given mass function of $dn/dm \propto m^{-\gamma}$ 
each obeying a universal functional form for $f$. The volume
distribution function can then be written as 
\be
v(f)=\int_{0}^{\mu M} \frac{dn}{dm}v_m(f)dm \propto f^{-(4-\gamma)}
\label{eq:super}
\ee
where $\mu M$ is the mass of the largest sub halo. However, 
for $\gamma=1.9$, as derived by \citep{2004MNRAS.348..333D}, this model predicts
$v(f) \propto f^{-2.1}$, rather than $v(f) \propto f^{-2.5}$ as found
in \citet{arad04}. \citet{asca05} modified this model by pointing  out
that the lower limit of the  integral in \equ{super} cannot be zero (model
AB) since the resolution of the simulation imposes a limit on the
minimum mass that a sub halo can have. For a
halo sampled with a finite number of particles each of mass $m_p$ 
the minimum mass of a sub halo is $m_{min} \sim 100 m_p$. 
The analysis as done in \citet{asca05}
assumes the sub halos to be Hernquist spheres and 
approximates its distribution function by a double power law
\be
v_m(f) =  
\left \{
\begin{array}{ll}
 5.46 \times 10^{-38} m^3(\frac{f}{k/m})^{-1.56} \ \ f \leq
   k/m \ \ \ \ \ \\
 5.46 \times 10^{-38} m^3(\frac{f}{k/m})^{-2.80} \ \ f \geq
   k/m \ \ \ \ \ 
\end{array}
\right.
\ee  
where $k=3.25 \times 10^{18} \Msun^{2}\Mpc^{-3}(\kms)^{-3}$.
The distribution function can then be written as 
\be
v(f) & =  &
3.18(\frac{f}{k})^{-2.1}-\frac{m_{min}^{0.54}}{0.54}(\frac{f}{k})^{-1.56}
-\frac{m_{min}^{-0.7}}{0.7}(\frac{f}{k})^{-2.8}   \ \ \ \ \ 
\label{eq:abmodel}
\ee
for $k/m_{max} \le f \le k/m_{min}$ and  
\be
v(f) \propto 
\left \{
\begin{array}{ll}
f^{-1.56} \ \ f \le k/m_{max} \ \ \ \ \ \ \ \ \ \ \ \ \\
f^{-2.80} \ \ f \ge k/m_{min} \ \ \ \ \ \ \ \ \ \ \ \ 
\end{array}
\right.
\ee
In \fig{alf1} we plot the slope parameter $\alpha$ as function of $f$
as predicted by the AD and AB Toy models (\equ{abmodel}). It can be seen that
in the limit the parameter $m_{min} \rightarrow 0$ and parameter
$m_{max} \rightarrow \infty$ the AB model approaches the AD model. 
We can see that either model fails to reproduce the behavior seen
in simulations.

\begin{figure}
    \centering \includegraphics[width=0.45\textwidth]{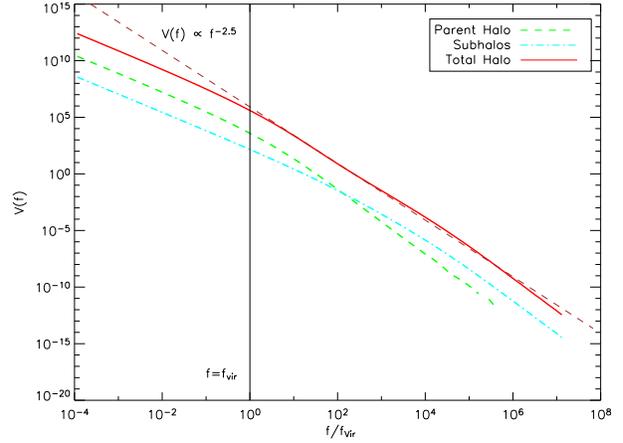}
  \caption[ Volume distribution function of phase space density,
    $v(f)$ as predicted by the parent $+$ substructure model proposed here.]{ Volume distribution function of phase space density,
    $v(f)$ as predicted by the parent $+$ substructure model proposed here. Curves
    for the parent halo and the sub halos were shifted vertically by
    two decades for clarity. The model $v(f)$ shows a slight hump
    beyond $f/f_{Vir}=10^2$ as compared to the constant slope 
    $v(f) \propto f^{-2.5}$ behavior. This is the point where the
    subhalo's contribution to $v(f)$ starts to dominate over the parent halo's
    contribution.  
\label{fig:vff_s}}
\end{figure}

\begin{figure}
    \centering \includegraphics[width=0.45\textwidth]{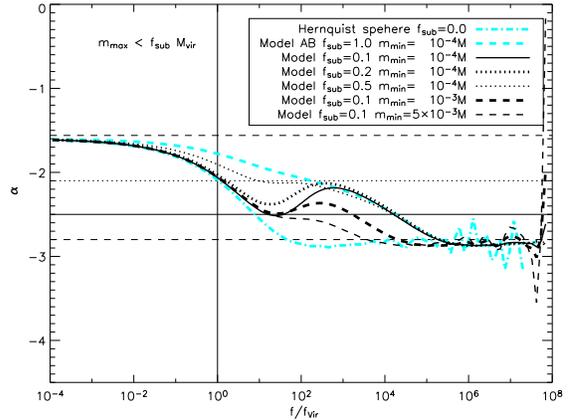}
  \caption[Dependence of slope parameter $\alpha$ on $f$ as
    predicted by the  parent $+$ substructure model proposed here.]{ Dependence of slope parameter $\alpha$ on $f$ as
    predicted by the  parent $+$ substructure model proposed here for
    different values of the parameters, sub halo mass fraction
    $f_{sub}$ and minimum mass of
    sub halo $m_{min}$. The profile has a minimum at $\log(f/f_{Vir})
    \sim 1.5$ and maximum at  $\log(f/f_{Vir}) \sim 3$. As $f_{sub}$
    increases (keeping $m_{min}=10^{-4} M$ constant) the minimum point
    of $\alpha$ moves up till it matches
    with with the $f_{sub}=1$ AB Toy model. On the other hand as
    $m_{min}$ is increased (keeping $f_{sub}=0.1$ constant) the
    maximum point of $\alpha$ drops down and ultimately it merges with
    the substructure-less Hernquist profile $f_{sub}=0$.
\label{fig:alf_s2}}
\end{figure}

In both the models it was assumed that the entire halo is made up by
superposition of sub halos with a mass function given by $dn/dm \propto m^{-\gamma}$.
In the analysis done by  \citet{2004MNRAS.348..333D}, where this mass
function was determined, the background parent halo which,  which
accounts about $90\%$ 
of the total mass, is excluded from the calculation.
The parent halo here is not a part of the substructure population.
We take this fact into account and develop a 
model in which we account separately for the contribution 
of the parent halo.
The halo consists of 1) the parent halo with mass
$(1-f_{sub})M$ modeled as a Hernquist sphere and 2) the substructure
of total mass $f_{sub}M$ which is modeled as a superposition of 
Hernquist spheres with a mass function of $dn/dm \propto m^{-\gamma}$.
To calculate the scale radius $a$ of a sub halo of mass $m$ 
we use the virial scaling relation $M_{Vir} \propto R_{Vir}^3$
which gives $m \propto a^3$  (assuming concentration parameter to be
same for all sub halos). 
In \fig{vff_s} we plot the volume distribution function as predicted
by this model for $f_{sub}=0.1, m_{min}=10^{-4}M$. 
In order to calculate $v(f)$ we employ a semi-analytic 
technique. We generate a sub halo population corresponding to the
given mass function and mass fraction $f_{sub}$ and then for a given
value of $f$ we sum the volume contribution of each sub halo along
with the parent halo
to give the total $v(f)$. The $v(f)$ for each sub halo is determined
using \equ{hernf} and \equ{hernv}. 
The total $v(f)$ as predicted by this model is close to the expected 
$v(f) \propto f^{-2.5}$ behavior but there is a presence of a slight hump in
the high $f$ part. This is similar to what we saw for $\Lambda$CDM
halos \fig{vff_r}.
In the high $f$ part $v(f)$ is dominated by
the substructure component the transition being at around
$f/f_{Vir}=10^2$. In \fig{alf_s2} we plot the slope parameter $\alpha$
as predicted by the model for various values of $f_{sub}$ and
$m_{min}$.

\begin{figure}
    \centering \includegraphics[width=0.45\textwidth]{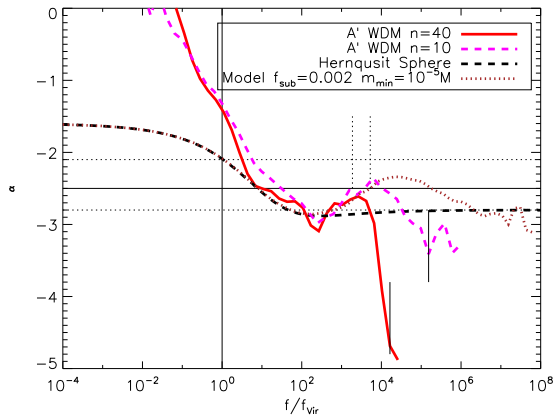}
   \caption{Effect of changing the number of smoothing neighbors on the 
     slope parameter $\alpha$ for a WDM halo. Results are shown for kernel 
     smoothing with smoothing neighbors $n=40$ and $n=10$. The slope parameter 
     $\alpha$ for a Hernquist sphere and a model with $m_{min}=10^{-5}$ and 
     $f_{sub}=0.002$ is also plotted alongside.
\label{fig:alf_wdm3}}
\end{figure}

\begin{figure}
    \centering \includegraphics[width=0.45\textwidth]{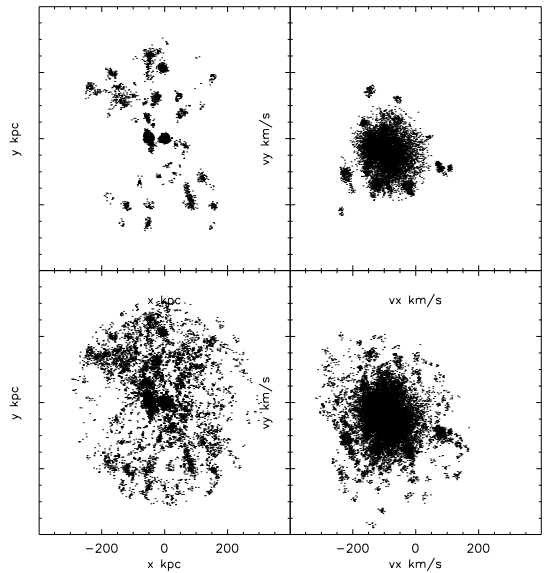}
  \caption{ The x Vs y and $V_x$ Vs $V_y$ scatter plot of particles  having 
    phase space density above $10^3 f_{Vir}$ for a warm dark matter halo.
    In top panels the density is evaluated  by using kernel smoothing 
    with 40 smoothing neighbors while in lower panels the density is 
    evaluated using 10 smoothing neighbors.
\label{fig:warmxvx}} 
\end{figure} 

In \fig{alf_r} $\alpha(f)$ corresponding to 
model with parameters $f_{sub}=0.05, m_{min}=10^{-5}M$ is compared 
against $\alpha(f)$ for simulated  halos. It can be seen that
the model (analytical profile) is successful in qualitatively 
explaining the behavior of the simulated halos 
(namely the dip and the peak) but there is still some difference 
at the low $f$ end.
At the low $f$ end near $f_{Vir}$, parameter $\alpha$ rises much more
sharply as compared to the model, even after
taking the truncation effect into account.

In (\fig{alf_wdm3}) we show the effect of varying the number of 
smoothing neighbors on the $\alpha$ profile of a WDM halo.
Lowering the number of smoothing neighbors to $10$ makes the slope 
parameter rise to a peak at the high $f$ end. 
Since with $n=10$ dispersion in density estimates is high, 
this also results in a slight flattening of the $\alpha$ profile 
around $f=f_{Vir}$, where $\alpha$ is found to rise steeply. 
Plotted alongside is the $\alpha$ profile of the best fit 
parent $+$ sub halo model.
The $\alpha$ profile of the WDM halo is consistent with a  
model having substructure mass fraction $f_{sub}=0.002$.
In \fig{warmxvx} we plot the particles having $f/f_{Vir}>10^3$
in both real and velocity space. In top panels the density 
was estimated using $n=40$ neighbors, while for 
lower panels the density was estimated using $n=10$ neighbors.
It can be seen from the figure that warm dark matter halo is not 
completely free from substructure. More substructure is resolved using 
smaller number of smoothing neighbors. The fact that even such a small 
amount of substructure can be detected demonstrates the superior 
ability of the estimator in resolving the high density regions.
It also suggests that the slope parameter $\alpha$ plotted as a 
function of $f$ can be used as a sensitive tool to estimate the amount 
of substructure and and the mass function of sub halos.

\begin{figure}
    \centering \includegraphics[width=0.45\textwidth]{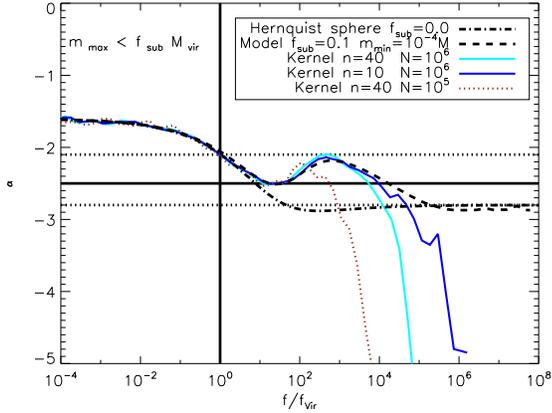}
  \caption[Dependence of slope parameter $\alpha$ on $f$ as
    recovered by EnBiD from a parent $+$ substructure
    model with $f_{sub}=0.1$
    and $m_{min}=10^{-4}M$.]{ Dependence of slope parameter $\alpha$ on $f$ as
    recovered by EnBiD from a parent $+$ substructure
    model. The fraction of mass in the form of substructure  is $f_{sub}=0.1$,  
    and the minimum mass of the substructure is $m_{min}=10^{-4}M$, $M$ being
    the total mass of the system. The theoretically 
    expected slope parameter for the above model and for a Hernquist 
    sphere without any substructure is also plotted alongside.
    For a system sampled with $10^6$ particles, the parameter $\alpha$ can 
    be accurately predicted till $f/f_{Vir}=10^4$ using kernel smoothing 
    with $10$ smoothing neighbors.
\label{fig:alf_s3}}
\end{figure}

To further check the efficiency of the code in reproducing the phase
space density of a system with substructure we generated a mock system with
$f_{sub}=0.1$ and $m_{min}=10^{-4}$, and calculated its phase space
density using EnBiD. The results are shown in \fig{alf_s3} .
The sub halos where distributed uniformly inside the virial radius of
the parent halo and their center of mass velocity was also chosen so
as to have a uniform random distribution within a sphere of radius
$V_{Vir}$ in velocity space. For a system modeled with $10^6$ particles, 
the phase space structure till $f=10^4 f_{Vir}$ is successfully reproduced 
by using kernel smoothing with $10$ smoothing neighbors.
If $40$ smoothing neighbors are used the high density regions are
poorly resolved.  Lowering the total number of particles in the system 
also leads to poor resolution at the high $f$ end.

\section{Discussion \& Conclusions}

We have presented a method for estimation of densities in a multi-dimensional 
space based on binary space partitioning trees \citep{asca05}. We implement a node splitting criterion that uses 
Shannon Entropy as a measure of information available in a particular 
dimension. The new algorithm makes the scheme metric free and 
recovers maximum information available from 
the data with a minimum loss of resolution. In our tests
on systems whose density distribution is known analytically, 
we find significant improvement in estimated densities 
as compared to earlier algorithms.

We suggest how kernel-based schemes (SPH) or in general any
metric based scheme can be implemented within the framework of 
the new algorithm: the algorithm EnBiD is used to determine the 
metric at any given point, which has the property that
locally the co-variance of the data points has a similar value 
along all dimensions. Next we incorporate this metric into
kernel-based schemes and use them for density estimation. 
We also show that SPH schemes suffer from a bias in their density
estimates. We suggest a prescription that can successfully correct the bias.

As an immediate application, we employ this method to analyze the 
phase space structure of dark matter halos obtained from N-Body
simulation with a $\Lambda$CDM cosmology. 
We find evidence for slight deviations from the near power law
behavior of the volume distribution function $v(f)$ of halos in such 
simulations. At the high $f$ end there is slight hump and the low $f$ 
end there is significant flattening. 
We also analyzed a $WDM$ halo and found that its slope parameter
profile  $\alpha(f)$ at the high $f$ end is consistent with that of 
an equilibrium Hernquist sphere having a very small amount of 
mass ($0.2\%$) in the form of substructure.

In $\Lambda$CDM halos the contribution to the volume distribution 
function at the high $f$ end is dominated by the presence of 
significant amount of substructure. We devise a toy model in which 
the halo is modeled as a Hernquist
sphere and the substructure is modeled as a superposition of Hernquist
spheres with a fixed mass fraction $f_{sub}$ and a mass function
$dn/dm \propto m^{-1.9}$.  We demonstrate that this reproduces 
the behavior of $v(f)$ as seen in simulations.

The behavior of $v(f)$ and $\alpha(f)$
depends upon the parameters $f_{sub}$,mass function  $dn/dm$ of sub halos, 
and $m_{min}$ the minimum mass of the sub halo.
Since the mass function of sub halos and their fraction 
$f_{sub}$ depends upon the power spectrum of initial conditions and on
the cosmology adopted, the phase
space structure of the halos might have an imprint of cosmology and 
initial conditions which might be visible in the profile $\alpha(f)$.

Although the simple toy model that we propose here 
can explain the basic properties of the volume
distribution function there is still some difference 
at the low $f$ end. The flattening at low $f$ end is more 
pronounced in simulated halos as compared to those of model halos, even after
taking the truncation effect into account. 
Further improvements on the model described include: 
The toy model assumes that all sub halos obey the same 
virial scaling relation while in simulation there should be slight
dependence on the time of formation of the sub halo. 
Moreover the  sub halos may be tidally truncated and stripped 
and so their density profile may be different from that of 
a pure Hernquist sphere \citep{2003ApJ...584..541H,2004ApJ...608..663K}. 
Furthermore, there might be a radial dependence 
on the properties of sub halos. A detailed model which takes into 
account these effects might help explaining the phase space properties 
more accurately.

The issue of universality in the behavior of the
volume distribution function still deserves further investigation.  
For the four halos that we have analyzed one of them had a nearly
flat $\alpha(f)$ profile and the others showed a characteristic dip
at $f \sim 10 f_{Vir}$ and a corresponding rise which peaks at around
$f \sim 10^3 f_{Vir}$. Larger samples of halos  
need  to be investigated in order to put these results on a sound 
statistical basis. The differences that are seen in the properties of halos
might be due to varying degree of virialization. 
The second concern is regarding the role of numerical 
resolution on the behavior of the volume distribution function.
In the model the shape of the $\alpha(f)$ profile
depends upon the minimum mass $m_{min}$ of the sub halo used to
model the sub halo population. According to the model  $\alpha(f)$ has a
minimum at around $f/f_{Vir} \sim 10$ and then it rises to a peak at
around $f/f_{Vir} \sim 10^3$ whose maximum
value is determined by the logarithmic slope of the mass function and 
is given be $-(4-\gamma)$. Beyond this point increasing the resolution
should make the  $\alpha(f)$ reach a plateau and then fall off once 
it reaches the resolution  limit of the simulation which occurs
approximately at $f_{relax}/f_{Vir} \sim 10^{-2}M_{Vir}/m_p$. 
This suggests that a proper convergence
study needs to be done to establish the universality in the phase space 
behavior of the halos.  At higher resolution existence of a 
behavior different from the toy model suggested here would imply 
that there are some physical processes at work which significantly 
alter the properties of low mass sub halos and drive the system towards 
a universal behavior e.g. the one with a constant slope.

Our analysis here shows that the phase space properties of the halos
that are roughly consistent with equilibrium
spherical models with a given density profile in real space.
A question of fundamental importance is regarding the origin of the
universal behavior of these density profiles as seen in simulations.
A clue to which might be found by studying as to how the system 
approaches equilibrium.
The evolution of the distribution function of 
collisionless particles is governed by the collisionless Boltzmann 
equation. Since the coarsely-grained 
distribution function of collisionless particles can be measured
directly with EnBiD, this offers interesting opportunities to study the processes of phase
mixing and violent relaxation, which help the system to reach equilibrium. 
It might be interesting in this context to study the evolution
of the volume distribution function of the halos with time.

Another interesting application of this method is to study the 
distribution function of equilibrium systems e.g. a disk that
hierarchically grows inside a halo. 
One can study the distribution function of these
systems and this can in turn be used to construct equilibrium models.

Finally we would like to point out a potential improvement in the code.
If the density distribution in any dimension is linearly 
independent of the other dimensions then this offers an opportunity
to further improve the density estimates by measuring the density 
distributions in different dimensions separately. The concept of
mutual information offers one such way to quantify this linear
dependence or independence. An algorithm can be developed which 
can exploit this feature and improve the density estimates in
situations where the data offers such an opportunity.

\section*{Acknowledgments}
We are grateful Vince Eke for his help
with the initial conditions of the simulations and Stefan Gottloeber
for providing one of his earlier simulations. This
work has been supported by grants from the U.S. National Aeronautics
and Space Administration (NAG 5-10827), the David and Lucile Packard
Foundation.

 \bibliographystyle{mn2e}
 \bibliography{../../LaTeX/BibTeX/DATABASE,../../LaTeX/BibTeX/PREPRINTS}

\begin{thebibliography}{ }


\bibitem[Arad et al.(2004)]{arad04} Arad, I., Dekel, A.,  Klypin, A.\ 2004, \mnras, 353, 15 
\bibitem[Ascasibar \& Binney(2005)]{asca05} Ascasibar, Y.,  Binney, J.\ 2005, \mnras, 356, 872 
\bibitem[Bernardeau \& van de Weygaert(1996)]{1996MNRAS.279..693B} Bernardeau, F., \& van de Weygaert, R.\ 1996, \mnras, 279, 693 
\bibitem[Binney \& Tremaine (1987)]{Binney87} Binney, J.,   Tremaine S., 1987, Galactic Dynamics (Princeton Univ Press)
\bibitem[Binney(2004)]{2004MNRAS.350..939B} Binney, J.\ 2004, \mnras, 350, 939 
\bibitem[Couchman(1991)]{1991ApJ...368L..23C} Couchman, H.~M.~P.\ 1991, \apjl, 368, L23 
\bibitem[De Lucia et al.(2004)]{2004MNRAS.348..333D} De Lucia, G., Kauffmann, G., Springel, V., White, S.~D.~M., Lanzoni, B., Stoehr, F., Tormen, G., \& Yoshida, N.\ 2004, \mnras, 348, 333 
\bibitem[Diemand et al.(2004)]{2004MNRAS.348..977D} Diemand, J., Moore, B., Stadel, J., \& Kazantzidis, S.\ 2004, \mnras, 348, 977
\bibitem[Gershenfeld (1999)]{Neil99} Gershenfeld N., \ 1999, {The Nature of Mathematical  Modeling}. Cambridge University Press
\bibitem[Gingold \& Monaghan(1977)]{1977MNRAS.181..375G} Gingold, R.~A., \& Monaghan, J.~J.\ 1977, \mnras, 181, 375 
\bibitem[Hayashi et al.(2003)]{2003ApJ...584..541H} Hayashi, E., Navarro, J.~F., Taylor, J.~E., Stadel, J., \& Quinn, T.\ 2003, \apj, 584, 541 
\bibitem[Hernquist(1990)]{1990ApJ...356..359H} Hernquist, L.\ 1990, \apj, 356, 359
\bibitem[Hockney \& Eastwood(1981)]{HE} Hockney, R. ~W., \& Eastwood,J. ~W. \ 1981,{\em Computer simulations using particles}, (New York: McGraw-Hill)
\bibitem[Kazantzidis et al.(2004)]{2004ApJ...608..663K} Kazantzidis, S., Mayer, L., Mastropietro, C., Diemand, J., Stadel, J., \& Moore, B.\ 2004, \apj, 608, 663 
\bibitem[Kravtsov et al.(1997)]{1997ApJS..111...73K} Kravtsov, A.~V., Klypin, A.~A., \& Khokhlov, A.~M.\ 1997, \apjs, 111, 73 
\bibitem[Loftsgaarden \& Quesenberry(1965)]{lq65} Loftsgaarden D. ~O., and Quesenberry, C.~P., \ 1948 ,A non  parametric estimate of a multivariate density function. {\em Annal. Math. Statist.}, 36:1049-1051.
\bibitem[Lucy(1977)]{1977AJ.....82.1013L} Lucy, L.~B.\ 1977, \aj, 82, 1013 
\bibitem[MacKay (2003)]{Kay03} MacKay D. ~J. ~C.,\ 2003, {Information Theory,Inference, and Learning Algorithms}. Cambridge University Press
\bibitem[Melchiorri, Bode, Bahcall, \&  Silk(2003)]{2003ApJ...586L...1M} Melchiorri, A., Bode, P., Bahcall,  N.~A., \& Silk, J.\ 2003, \apjl, 586, L1
\bibitem[Monaghan \& Lattanzio(1985)]{1985A&A...149..135M} Monaghan, J.~J., \& Lattanzio, J.~C.\ 1985, \aap, 149, 135 
\bibitem[Monaghan(1992)]{1992ARA&A..30..543M} Monaghan, J.~J.\ 1992, \araa, 30, 543 
\bibitem[Navarro et al.(2004)]{2004MNRAS.349.1039N} Navarro, J.~F., et al.\ 2004, \mnras, 349, 1039 
\bibitem[Navarro, Frenk, \& White(1996)]{1996ApJ...462..563N} Navarro,  J.~F., Frenk, C.~S., \& White, S.~D.~M.\ 1996, \apj, 462, 563
\bibitem[Navarro, Frenk, \& White(1997)]{1997ApJ...490..493N} Navarro,  J.~F., Frenk, C.~S., \& White, S.~D.~M.\ 1997, \apj, 490, 493
\bibitem[Okabe et al.(1992)]{1992stca.book.....O} Okabe, A., Boots, B., \& Sugihara, K.\ 1992, Wiley Series in Probability and Mathematical Statistics, Chichester, New York: Wiley, 1992, 
\bibitem[Okabe(2000)]{2000stca.conf.....O} Okabe, A.\ 2000, Spatial tessellations : concepts and applications of voronoi diagrams.~2nd ed.~By Atsuyuki Okabe ...[et al] Chichester ; Toronto : John Wiley {\&} Sons, 2000., 
\bibitem[Press et al.(1992)]{1992nrca.book.....P} Press, W.~H., Teukolsky, S.~A., Vetterling, W.~T., \& Flannery, B.~P.\ 1992, Cambridge: University Press, |c1992, 2nd ed.,
\bibitem[Schaap \& van de Weygaert(2000)]{2000A&A...363L..29S} Schaap, W.~E., \& van de Weygaert, R.\ 2000, \aap, 363, L29  
\bibitem[Shannon (1948)]{Shan48} Shannon, C.~E., \ 1948 ,A mathematical theory  of communication. {\em Bell System Tech. J.}, 27:379--423.
\bibitem[Shannon (1949)]{Shan49} Shannon C.~E., \& Weaver W.,\ 1949, {The Mathematical Theory of Communication}.  University of Illinois Press.
\bibitem[Shapiro et al.(1996)]{1996ApJS..103..269S} Shapiro, P.~R., Martel, H., Villumsen, J.~V., \& Owen, J.~M.\ 1996, \apjs, 103, 269 
\bibitem[Silverman(1986)]{1986desd.book.....S} Silverman, B.~W.\ 1986, Monographs on Statistics and Applied Probability, London: Chapman and Hall, 1986,  
\bibitem[Spergel et al.(2003)]{2003ApJS..148..175S} Spergel, D.~N.~et  al.\ 2003, \apjs, 148, 175
\bibitem[Springel, Yoshida, \& White(2001)]{Springel01}  Springel, V., Yoshida, N., \& White, S.~D.~M.\ 2001, New Astronomy,  6, 79
\bibitem[Stadel(1995)]{Sta95} Stadel, J. \ 1995, \newline http://www-hpcc.astro.washington.edu/tools/smooth.html
\end{thebibliography}

\appendix
\section{Kernel Density Estimate}
For the so called kernel density estimate (KDE) a kernel $W$ is defined such that 
\begin{equation}
\int W({\bf x,h}) d^d {\bf x}=1
\end{equation}
The density estimate of a discretely set of N particles at a point ${\bf x}$ is
given by 
\begin{equation}
\rho({\bf x})=\sum_i  m_i W({ \bf x_i-x,h})
\end{equation}
while the probability density $\hat{f}({\bf x})$ is given by
\begin{equation}
\hat{f}({\bf x})=\frac{1}{N}\sum_i W({\bf x_i-x},{\bf h})
\label{eq:fx}
\end{equation}

The smoothing parameter ${\bf h}$ is chosen such that it encloses a fixed
number of neighbors $N_{smooth}$.
Assuming spherical symmetry the kernel can be written in terms of a 
radial co-ordinate $u$ only. Some of the popular choices are Gaussian
function and the B-splines \citep{1985A&A...149..135M}. The later is preferred due to its compact 
support. 
A $d$ dimensional multivariate bandwidth spherical kernel can be written as
\begin{equation}
W({\bf x},{\bf h})=\frac{ f_d W_d(u)}{\Pi_{i=1}^{d} h_i}
\end{equation}
where 
\begin{equation}
u = \sqrt{\sum_{i=1}^{d} \left( \frac{x_i}{h_i}\right)^2}
\end{equation}
and the normalization $f_d$ is given by
\begin{equation}
f_d = \frac{1}{\int_{0}^{1}W(u) S_d u^{d-1}du}
\label{eq:fd}
\end{equation}
$S_d$ being the surface of a unit hyper-sphere in $d$ dimensions $V_d$ its volume.
\begin{eqnarray}
S_d& =& 2 \pi^{d/2}/\Gamma(d/2) \textrm{\ \ ;\ \ $V_d=S_d/d$ }
\end{eqnarray}
Some popular kernels are given below  and their normalizations constants $f_d$
are listed in \tab{corr1}
\begin{eqnarray}
W_{Gaussian}(u)= \exp(-u^2) \textrm{\ \ ;\ } f_d=\frac{1}{\pi^{d/2}}
\end{eqnarray}

\begin{eqnarray}
W_{Top-Hat}(u)= \left\{
\begin{array}{ll}
 1 & \ \ 0 \le  u \leq 1 \\
 0 & \ \ \rm{otherwise}
\end{array}
\right. 
\textrm{\ \ ;\ } f_d=\frac{1}{V_d}
\end{eqnarray}

\begin{eqnarray}
W_{Spline}(u)= \left\{
\begin{array}{ll}
 1-6u^2+6u^3 & \ \ 0 \le  u \leq 0.5 \\
 2(1-u)^3 & \ \ 0.5 \le  u \leq 1 \\
 0 & \ \ \rm{otherwise}
\end{array}
\right. 
\end{eqnarray}

\begin{eqnarray}
W_{Epanechikov}(u)= \left\{
\begin{array}{ll}
 (1-u^2) & \ \ 0 \le  u \leq 1 \\
 0 & \ \ \rm{otherwise}
\end{array}
\right. 
\end{eqnarray}

\begin{eqnarray}
W_{Bi-Weight}(u)= \left\{
\begin{array}{ll}
 (1-u^2)^2 & \ \ 0 \le  u \leq 1 \\
 0 & \ \ \rm{otherwise}
\end{array}
\right. 
\end{eqnarray}
For kernels in product form 
\begin{equation}
W({\bf x},{\bf h})=\frac{ \Pi_{i=1}^{d} f_1 W(u_i)}{\Pi_{i=1}^{d} h_i}
\end{equation}
where $u_i=x_i/h_i$ and $f_1$ is the corresponding one dimensional 
normalization factor as given by \equ{fd}.

\begin{table}
  \caption  {Normalization constants for various dimensions}
\begin{center}
\begin{tabular}{@{}|c|ccc|}
\hline
Dimension & \multicolumn{3}{|c|}{Normalization $f_d$} \\
\cline{2-4}
  d & Spline & Epanechnikov & Bi-Weight \\
\hline
           1 &        1.3333369 &       0.75000113 &       0.93750176 \\
           2 &        1.8189136 &       0.63661975 &       0.95492964 \\
           3 &        2.5464790 &       0.59683102 &        1.0444543 \\
           4 &        3.6606359 &       0.60792705 &        1.2158542 \\
           5 &        5.4037953 &       0.66492015 &        1.4960706 \\
           6 &        8.1913803 &       0.77403670 &        1.9350925 \\
           7 &        12.748839 &       0.95242788 &        2.6191784 \\
           8 &        20.366416 &        1.2319173 &        3.6957561 \\
           9 &        33.380983 &        1.6674189 &        5.4191207 \\
          10 &        56.102186 &        2.3527875 &        8.2347774 \\
\hline
 \end{tabular}
\end{center}
  \label{tb:corr1}
\end{table}

\subsection{Optimum Choice of Smoothing Neighbors}
If $\hat{f(x)}$ is the estimated probability density of a field $f(x)$ then 
its {\it mean square error} (MSE) can be written in terms  of its bias 
$\beta(x)$ and variance $\sigma(x)$. Bias of an estimate is given by 
\be
\beta(x) &=&  \langle \hat{f}(x)\rangle- f(x)
\ee
while its variance is 
\be
\sigma^2(x) &=& \langle \left[\hat{f}(x)-\langle \hat{f}(x)\rangle \right]^2\rangle 
\ee
Hence mean square error is given by
\be
MSE [\hat{f}(x)] &= & \langle \left[\hat{f}(x)-f(x)\right]^2 \rangle \\
&=& \langle \left[\hat{f}(x)-\langle \hat{f}(x) \rangle + \langle \hat{f}(x)
  \rangle -f(x) \right]^2 \rangle \\
&=& \sigma^2(x)+\beta^2(x)
\ee
To get accurate estimates both bias and variance should be small.
Using the fact that
\be
\langle \sum_{i=1}^{N} A(x-x_i) \rangle = N\int A(x-x')f(x')dx'
\ee
the bias and variance of an estimator can be calculated by 
using \equ{fx} and expanding $f(x')$ as a Taylor series about $x$. 
For a $d$ dimensional multivariate kernel density estimate,the 
bias and variance are given by
\be
\beta(x) & \approx & \frac{h^2}{2} Tr[H_{f}(x)] \int u^2 W_d(u) S_d u^{d-1} du
\ee
where $H_{f}(x)=\frac{\partial^2 f}{\partial x_i \partial x_j}$ is the 
Hessian matrix of function $f(x)$.
\be
\sigma^2(x) & \approx & \frac{1}{n h^d} f(x) \int W_d^2(u) S_d u^{d-1} du \\
& \approx &  f^2(x)\frac{V_d}{N_{smooth}}\int W_d^2(u) S_d u^{d-1} du \\
& \approx &  f^2(x)\frac{V_d}{N_{smooth}}||W_d||^2_{2}\\
\label{eq:sigma}
\ee
$||W_d||^2_{2}$ being the $d$ dimensional $L^2$ norm of
kernel function $W_d(u)$.

\begin{figure}
    \centering \includegraphics[width=0.45\textwidth]{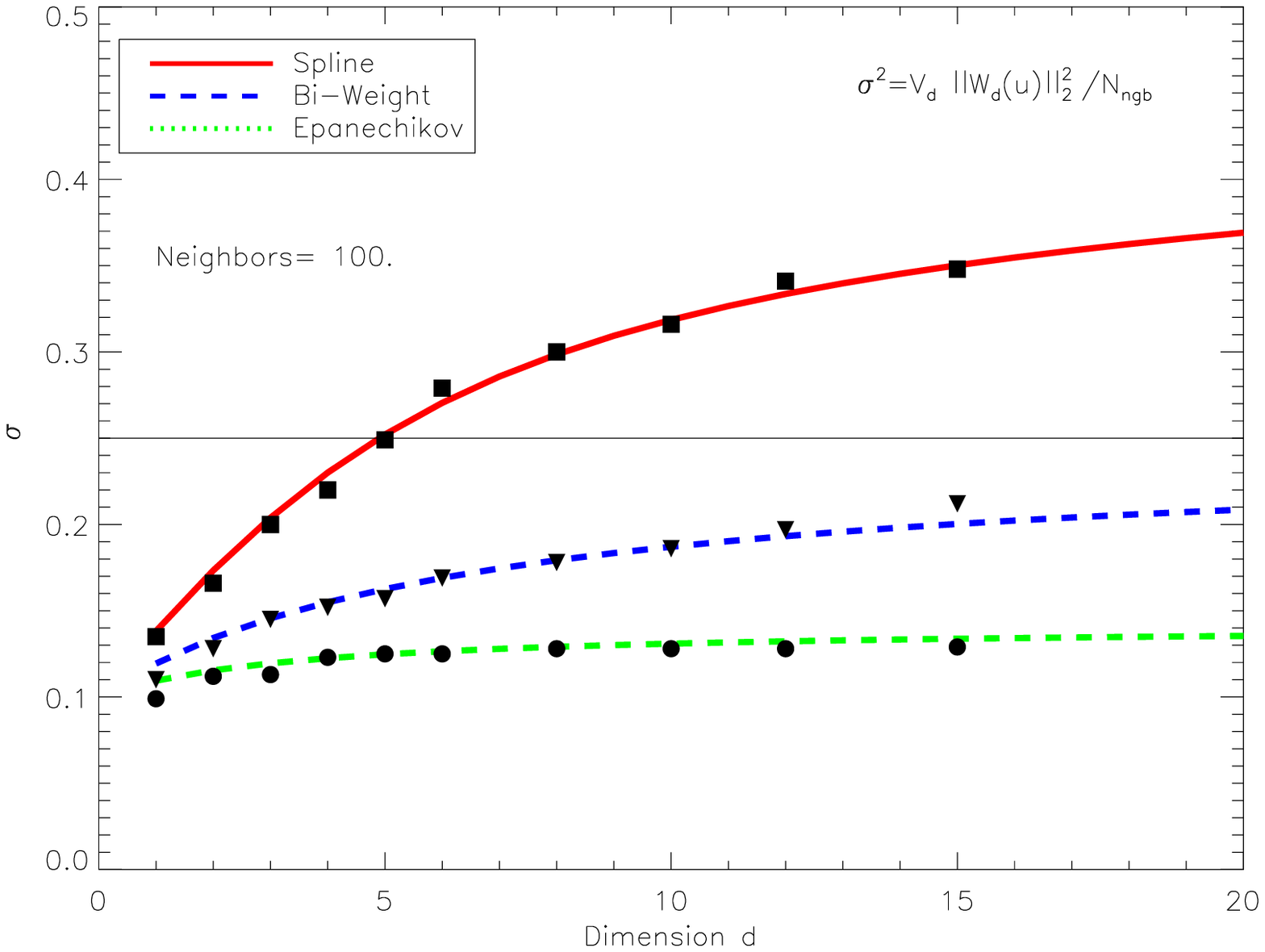}
  \caption{The variance of density estimates, 
    as obtained by kernel smoothing using 100 smoothing neighbors, as a function
    of number of dimensions.
The solid lines are calculated using \equ{sigma} while 
the points are the $\sigma$ extracted from a Poisson sampled 
data by applying kernel smoothing.
    \label{fig:sig_vs_dim}}
\end{figure}

Lowering $h$ or equivalently lowering $N_{smooth}$ lowers 
$\beta(x)$ but increases $\sigma(x)$. Ideally the optimum choice 
of $N_{smooth}$ is given by minimizing the MSE. The bias $\beta$, which depends 
on the second order derivative of the field, is small for slowly varying fields,
hence can be ignored. 
Since $\sigma(x) \propto 1/\sqrt{N_{smooth}}$, the variance
increases as $N_{smooth}$ is decreased. The minimum value of
$N_{smooth}$ that is needed to attain a given value of 
$\sigma(x)$ is the optimum choice of number of neighbors.
We define this lower limit on $\sigma$ as $0.25 f(x)$.  
In \fig{sig_vs_dim} $\sigma$ is plotted as a function 
of number of dimensions $d$ for $N_{smooth}=100$ (assuming $f(x)=1$). 
The variance as obtained by applying kernel smoothing on a Poisson sampled data 
with $N_{smooth}=100$ is also shown alongside. They are in agreement.
The variance $\sigma$ does not increase exponentially with number of dimensions.
Hence the optimum number of neighbors also do not  have to grow 
exponentially with the number of dimensions. This means that even 
in higher dimensions kernel smoothing can be efficiently done 
employing a small number of neighbors. 
In higher dimensions the efficiency of the nearest neighbor search 
algorithm is the main factor which determines the time required 
for kernel density estimation. 
It can also be seen from \fig{sig_vs_dim} that for a fixed number of 
neighbors the spline kernel gives maximum variance while the 
Epanechnikov kernel gives the lowest variance. 
\equ{sigma} can be used to calculate
the number smoothing neighbors $N_{smooth}$ required to achieve a given  
$\sigma$, for any given kernel in any arbitrary dimension. 
For density estimation with an Epanechnikov  kernel in $6$ dimensions,
$N_{smooth}=32$ gives a  variance of $\sigma=0.22$ which is 
equivalent to a variance of $0.1$ dex.

\subsection{Fraction of Boundary Particles}
For a system of $N$ particles uniformly distributed in a spherical region in 
a $d$ dimensional space the fraction of particles $f_b$ that lie on the 
boundary increases sharply with the number of dimensions $d$.
If $l$ is the mean inter-particle separation then $l=(V_dr^d/N)^{1/d}$
and the fraction $f_b$ is given by
\ben
f_b & = &(S_d r^2/l^2)/(V_dr^3/l^3)=dl/r=d(V_d/N)^{1/d} 
\een
For $N=10^6$, the fraction $f_b$ is $0.05$ and $0.79$ for $d=3$ and 
$d=6$ respectively.

\subsection{Anisotropic Kernels}
For planar structures which are not parallel to one of the 
co-ordinate axis one needs to adopt an anisotropic kernel to 
get accurate results. This is equivalent to a 
transformation with a rotation and a shear 
which diagonalizes the covariance matrix and then normalizes the 
eigenvalues \citep{1996ApJS..103..269S}. 
Let $H$ be a diagonal matrix such that $H_{ii}=h_i$ and ${\bf x'}=H^{-1}{\bf x}$.
If $C({\bf x'})$ is the covariance matrix locally
at point ${\bf x'}$ then the kernel is given by
\be
W({\bf x,h}) & = & \frac{f_d}{|D|^{1/2}|H|} W_d \left(|D^{-1/2}EH^{-1}{\bf x}|\right) 
\ee
where $E$ is the eigenvalue matrix that diagonalizes $C$ and $D$ is
the corresponding diagonal eigenvalue matrix. 
To keep the number of smoothing neighbors roughly constant we normalize the
eigenvalue matrix, $\bm{D}\rightarrow \bm{D}/|\bm{D}|^{1/d}$, this 
preserves the smoothing volume. To identify the neighbors 
that contribute to the density at $\bm{x}$ one now needs to select a
spherical region with radius ${\bf h'}={\bf h} \ \rm{max}(\bm{D^{1/2}})$.

\subsection{Bias in Spline Kernels}
Spline kernels have a bias in their estimated densities
i.e. they systematically overestimate the density. 
This is not present for a
regularly distributed data like a lattice or a glass like
configuration where the inter-particle separation is
constant \footnote{This bias does not affect the results in SPH 
simulations because the particles are not distributed randomly but
rather by the dynamics \citep{1992ARA&A..30..543M}. The dynamics of
the pressure forces results in a configuration which is 
regular and with nearly constant inter-particle 
separation.}. This only occurs for a data which has Poisson noise  
and whose density is measured at the location of the data points.
In some sense the bias is due to evaluation of the density at the
location of Poisson peaks in the density distribution.
The smaller the distance from the center the
greater the weight of the kernel. When the density is estimated at the
location of the particle the kernel assigns a very high weight to this
particle since its distance is zero. Below is shown a simple
calculation which demonstrates the bias in a spline kernel 
as compared to a top hat kernel which is free from such bias.
\be
\frac{f}{f_t} & = & \frac{\sum_{i=0}^{i=k} m W_i}{\rho_t} \\
& = & \frac{m W_{r=0} +\sum_{i=1}^{i=k} m W_i}{\rho_t} \\
\ee
Assuming that the top hat kernel gives the correct density 
$f_t=k/(V_dh^d)$. Taking one particle out from the smoothing 
region should roughly give a density of  
$\sum_{i=1}^{i=k} m W_i=m(k-1)/V_dh^d$.
\be
& = & \frac{m W_{r=0} +(k-1)m/(V_dh^d)}{km/(V_dh^d)} \\
& = & 1+\frac{f_d V_d-1}{k} \\
\label{eq:bias}
\ee

It can be seen from \equ{bias} that the bias decreases when the number 
of smoothing neighbors $k$ is increased. 
This bias can be removed by displacing the central particle
having $r=0$ to $r=hd/(1+d)$, $h$ being the radius of the smoothing sphere,
and $d$ the dimensionality of the space. 
This corresponds to the mean value of radius $r$ of a homogeneous sphere in a 
$d$ dimensional space . This correction should only be applied if the distribution of data is known to be irregular.  

\begin{figure}
    \centering \includegraphics[width=0.45\textwidth]{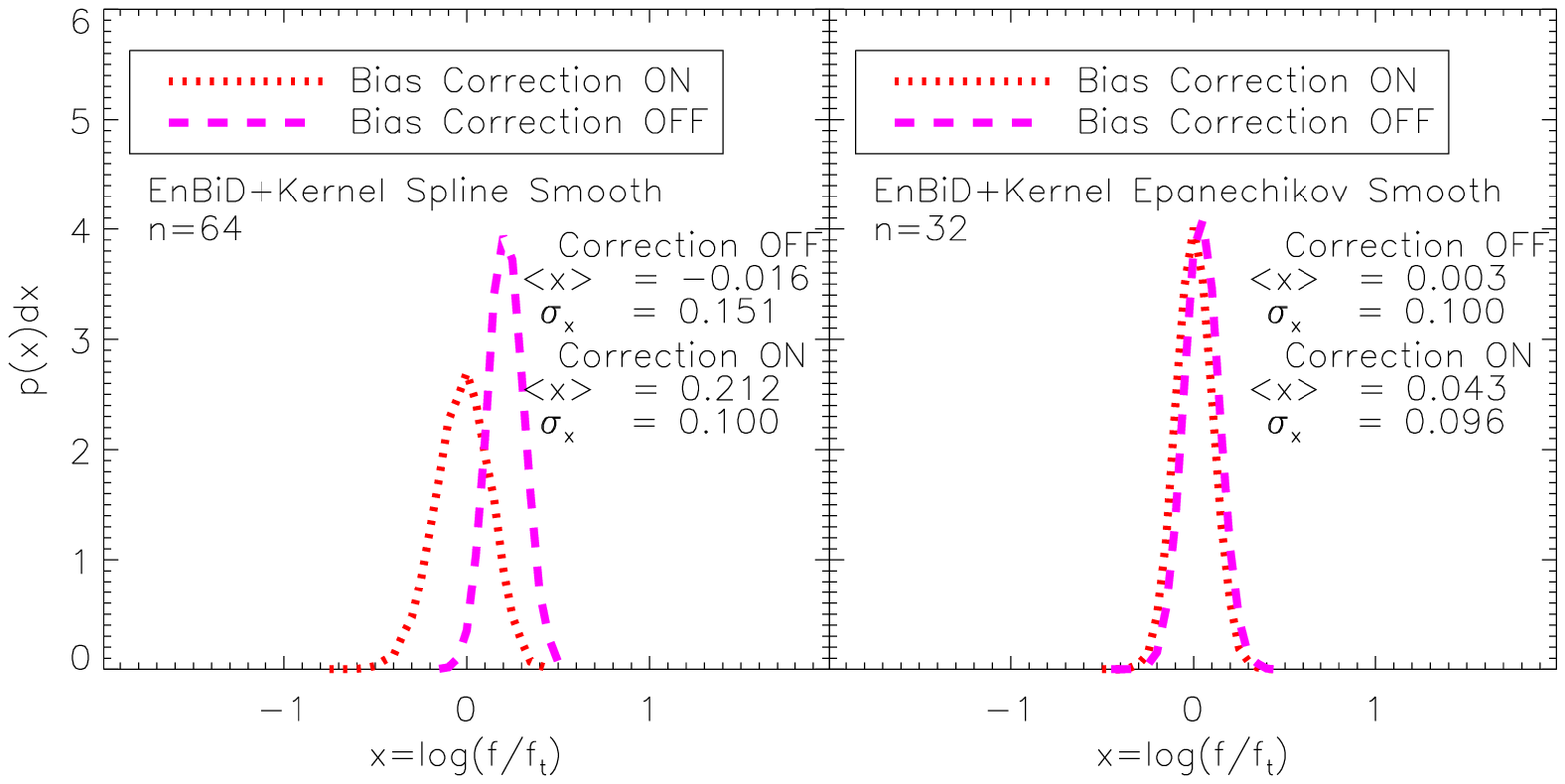}
  \caption{Kernel density estimates with and without bias correction.
    are shown for a system of $N=10^5$ particles distributed uniformly in a 6 dimensional space 
    with periodic boundaries.  Probability distribution $P(\log(f/f_t))$ is
    plotted for spline kernel with smoothing
    neighbors $n=64$ (left panel ) and Epanechnikov kernel 
    with $n=32$ (right panel). 
    The mean $<x>$ and  dispersion $\sigma_x$ of the best fit 
    Gaussian distribution to $x=\log(f/f_t)$, is also shown alongside.
\label{fig:bias_corr}}
\end{figure}

In \fig{bias_corr} kernel density estimates with 
and without bias correction  are shown for a system of $N=10^5$ particles distributed uniformly in a 6 dimensional space with periodic boundaries.  
In left panel the probability distribution $P(\log(f/f_t))$ is
plotted with and without bias correction, for kernel density estimate 
obtained using a spline function and smoothing neighbors $n=64$.
In right panel the probability distributions are
plotted for kernel density estimates obtained using an Epanechnikov 
function and smoothing neighbors $n=32$.
The bias given by mean $<x>$ of the best fit 
Gaussian distribution is also plotted alongside. 
According to \equ{bias}, in a $6$ dimensional space for spline kernels 
with neighbors $k=64$ the bias is $<\log(f_{sp}/f_t)>=0.21$ and 
for Epanechnikov kernel with $k=32$ the bias is $<\log(f_{Ep}/f_t)>=0.04$. 
These values are close to those shown in \fig{bias_corr} 
for uncorrected estimates.
The Epanechnikov kernel function has less bias than the 
spline kernel function.
After correction, for both the kernels, the bias is considerably reduced.

\end{document}